# The Critical Current Density of SNS Josephson Junctions and Polycrystalline Superconductors in High Magnetic Fields

A. I. Blair and D. P. Hampshire

*Superconductivity Group, Centre for Materials Physics,*
*Department of Physics, Durham University, DH1 3LE, UK*[*]

## Abstract

We investigate the in-field critical current density $J_c(B)$ of SNS Josephson junctions (JJs) and polycrystalline superconducting systems with grain boundaries modelled as Josephson-type planar defects, both analytically and through computational time-dependent Ginzburg–Landau (TDGL) simulations in 2D and 3D. For very narrow SNS JJs, we derive analytic expressions for $J_c(B)$ that to our knowledge are the first high field solutions for $J_c(B)$ for JJs across the entire applied field range up to the effective upper critical field $B_{c2}^*$. They generalise the well-known (low-field) exponential junction thickness dependence for $J_c$ from de Gennes, often used in the Josephson relation. We extend the new analytic expressions to describe wider junctions and confirm their validity using the TDGL simulations. These new results are then compared to the current densities found in superconductors optimised for high field applications. They provide an explanation for the $B^{-0.6}$ field dependence found for $J_c(B)$ in high temperature superconductors, and the Kramer field dependence and inverse power-law grain size widely found in many low temperature superconductors.

## I. INTRODUCTION

Probably the most important challenge in high field superconductivity is to understand and control the critical current density ($J_c$) of superconducting materials in high magnetic fields. The enormous dissipationless currents that technological superconducting materials can carry have made them essential components in large-scale high-field magnet systems, such as those used for high resolution NMR or to confine fusion plasmas. However in high magnetic fields, critical current densities in state-of-the-art commercial polycrystalline materials are still 2-4 orders of magnitude below theoretical (depairing) limits, suggesting significant potential for improvement and material optimisation [1]. Work on flux pinning over the decades has provided the qualitative understanding that $J_c$ can be increased in a superconductor through the introduction of pinning sites, which restrict the dissipative motion of vortices through the superconductor. However, a more quantitative description in these materials is limited by our understanding of the so-called 'grand summation problem': the problem of how the local vortex-vortex and vortex-pin interactions should be summed in order to obtain the macroscopic average $J_c$. For example, the proportion of vortices that are pinned at pinning sites, or how vortices relax after being depinned, remains unknown. Without such knowledge, our understanding of the vortex pinning and $J_c$ remains qualitative at best. Previous work has proposed the use of Josephson junctions (JJs) as analogues of grain boundaries for the basis of descriptions of flux flow and pinning in polycrystalline materials, computationally [2], and experimentally in both low and high temperature superconductors [3–5]. However, although some high field approximations have been proposed for very narrow junctions that lack vortices in the junction region [6, 7], there are no detailed analytic expressions for $J_c$ for any width of JJ in high fields up to the effective upper critical field $B_{c2}^*$ which address the complexity of vortices entering the superconducting electrodes [8, 9]. In this work we provide new analytic equations to describe very narrow JJs in high fields up to $B_{c2}^*$. We then extend the work to consider narrow junctions with many vortices both inside the junctions and in the superconducting electrodes. Our approach is to derive expressions for $J_c$ and confirm them to be valid using TDGL simulations. Necessarily, this work solves the 'grand summation' problem within the critical Josephson junction region itself, by including the non-uniform distribution of vortices in the junctions at $J_c$ [8, 10]. We also present 3D TDGL simulations of polycrystalline equiaxed superconducting systems, using superconducting-normal-superconducting (SNS) JJ structures for grain boundary regions. Comparison with $J_c$ in both LTS and HTS materials show that the TDGL simulations and equations derived here describe the well-known grain size and field dependence of

---

* Contact: d.p.hampshire@durham.ac.uk at Durham University, or alexander.blair@ukaea.uk who is now contactable at United Kingdom Atomic Energy Authority, Culham Science Centre, Abingdon, Oxfordshire, UK

the critical current in these materials and provide a new framework for their future development.

All large scale superconducting materials are polycrystalline, and contain a range of non-superconducting inclusions such as grain boundaries that are a barrier to current flow. Nevertheless, as predicted by Josephson [11], an electric current can flow across the barrier without the onset of dissipation since Cooper pairs are able to tunnel through these 'normal' regions. Optimised planar defects in superconducting systems must provide sufficient pinning to impede flux motion, without significantly limiting the current density that can flow through them [1]. In low applied magnetic fields $B_{\text{app}}$, the behaviour of $J_c(B_{\text{app}})$ of superconductor-normal-superconductor (SNS) junctions in tunnel-like geometries is very well known across a wide range of junctions widths. In very narrow systems, and in other geometries in which the current density is constrained to flow in one direction only, the critical current density that can flow across a thin normal metal sandwiched between two superconducting electrodes depends exponentially on the thickness of the normal barrier in the junction [12–14]. The local current density in this case is a sinusoidal function of the phase difference between the two superconducting electrodes [13–16]. When the dimensions of the superconductors become comparable to the magnetic penetration depth $\lambda_s$, the phase difference across the junction varies along the junction, and the (average) critical current density of the junction becomes highly dependent on the system geometry, with $J_c(B_{\text{app}}) \sim B_{\text{app}}^{-1/2}$ for well separated junctions in the thin film limit [8, 10]. For systems larger than $\lambda_s$ but smaller than the Josephson penetration depth $\lambda_J$, in which self-field effects in the junction region can be neglected, the gauge invariant phase difference across the junction varies approximately linearly with position along the junction (except near the junction edges), leading to the well-known Fraunhöfer-like dependence of $J_c$ on applied magnetic field [17, 18]. For very wide systems, in which the self field associated with the transport current through the junction is comparable to the applied magnetic field, the current-phase relation is multivalued depending on the number of vortices on the junction, and the critical current density depends on the magnetic history of the system [19]. Larger scale networks of SNS junctions have also been used in analytic models for the critical current of polycrystalline superconductors [3, 20]. However, these canonical low-field descriptions of SNS Josephson junctions have not been developed to consider fields up to the effective upper critical field of the superconducting regions, when the superconductors are in the mixed state and the order parameter is heavily suppressed below its Meissner state value. The presence of vortices in the surrounding superconductor strongly affects the critical current density that the junction can carry [21] when vortices are within the magnetic penetration depth $\lambda_s$ of the junction. Because there are no analytic solutions for the critical current of junction systems in this high field regime where $J_c$ decreases to zero and the superconductivity is destroyed in the electrodes, the detailed microscopic descriptions of JJs derived for low field applications have only had a very limited impact on the community trying to optimise grain boundaries that are pinning or limiting $J_c$ in high field superconductors. In this work, we try to further develop the interrelationship between the low-field JJ community and the high-field flux pinning community in applied superconductivity.

In this paper we find new solutions for the critical current density of narrow SNS junctions in all applied magnetic fields up to the upper critical field of the system, by extending the approach of Fink used in low fields [14] and developing the methodology of [22, 23] to account for the suppression of superconductivity in the superconducting electrodes in high fields. We verify our solutions against simulations based on time-dependent Ginzburg–Landau (TDGL) theory. TDGL theory has been used model the critical current density as a function of applied field for superconducting systems containing normal inclusions before [2, 24–26]. Whilst the conditions in which TDGL theory may be derived analytically from microscopic theory are formally quite narrow, restricted to the case of gapless superconductors limited by paramagnetic impurities or in the dirty limit, it has been widely applied in a phenomenological manner since it remains a relatively simple mathematical framework compared to microscopic theory in which the dynamics of vortex-vortex interactions are included self-consistently with Maxwell's equations for the evolution of electromagnetic fields in superconducting systems [27]. Indeed TDGL simulations of superconductors containing a periodic series of 'weak link' junctions in which the local temperature $T = T_c$ inside the junctions have previously been carried out [28], where the focus was on the vortex structure and dynamics through the junction, rather than how the junction properties affect $J_c(B_{app})$ as here. Next, we extend these very narrow width results to wider JJs, up to the scale of $\lambda_s$, to give in-field critical current densities. We demonstrate and discuss the qualitative agreement with the widely observed experimental results for $J_c(B_{app})$, namely the Kramer dependence [29] for low temperature superconductors such as Nb$_3$Sn [30, 31], Nb$_3$Al [32] and PbMo$_6$S$_8$ [33], and the power law dependence, $J_c(B_{app}) \sim B_{app}^{-0.6}$ for high temperature superconductors such as Bi$_2$Sr$_2$Ca$_2$Cu$_3$O$_x$ [34] and REBa$_2$Cu$_3$O$_7$ [5].

We shall first outline the computational method used to obtain critical current density as a function of applied field. We next validate our computational codes against the canonical low-field expressions for the critical current density of junctions, and find good agreement between our simulation and existing theory. We then present our new analytic solutions for the critical current density of very narrow junctions in all fields when the junction and superconductors are vortex-free and compare them to simulation. We then propose an extension of these very narrow junction results to physically relevant narrow system sizes, where there are fluxons both in the junctions and in the electrodes, and compare them to our simulations and relevant experimental systems. Finally, we present 3D TDGL simulations and visualisations of equiaxed polycrystalline systems with grain boundaries that are SNS Josephson junctions, which are consistent with the analytic forms presented and display optimised critical current densities that qualitatively display grain size and field dependencies that are widely observed in superconducting materials optimised for high field applications.

## II. TIME-DEPENDENT GINZBURG–LANDAU (TDGL) THEORY

In this work, we analyze Josephson junction systems entirely within the framework of the TDGL equations for gapless s-wave superconductors in the dirty limit [27]. Whilst the TDGL equations can be extended to consider weakly gapped superconductors or those with different order parameter symmetry (e.g. $d$-wave), we shall solve the TDGL equations in their simplest form for computational efficiency; nevertheless, we expect our results to hold qualitatively for a wide range of superconducting systems, as vortex-vortex interactions are included self-consistently with Maxwell's equations within the gapless s-wave TDGL equations. These normalised TDGL equations can be written as [35, 36]

$$\eta \left( \partial_t + \imath \mu \right) \psi = \left[ \sum_i (\partial_i - \imath A_i) m_i^{-1}(\boldsymbol{r}) (\partial_i - \imath A_i) + \alpha(\boldsymbol{r}) - \beta(\boldsymbol{r}) |\psi|^2 \right] \psi, \qquad (1)$$

$$\partial_t A_i + \partial_i \mu = -\kappa^2 m_i(\boldsymbol{r}) \left( \boldsymbol{\nabla} \times \boldsymbol{\nabla} \times \boldsymbol{A} \right)_i + \mathrm{Im} \left[ \psi^*(\partial_i - \imath A_i)\psi \right], \qquad (2)$$

where $\imath = \sqrt{-1}$ is the imaginary unit; $\kappa$ is the Ginzburg–Landau parameter of the superconductor; the index $i$ runs over the coordinate dimensions $x, y$ and $z$, which are normalised in units of the temperature-dependent zero-field coherence length $\xi_s$ of the superconductor in the $x$ direction; $t$ is time, normalised in units of a characteristic timescale $\tau = \mu_0 \kappa^2 \xi_s^2 / \rho_s$ where $\mu_0$ is the permeability of free space and $\rho_s$ is the normal state resistivity of the superconductor; $\psi$ is the order parameter, normalised in units of the bulk temperature dependent Meissner state value $\psi_0$; $\boldsymbol{A}$ is the magnetic vector potential, normalised in units of $\phi_0 / 2\pi \xi_s$, where $\phi_0$ is the magnetic flux quantum; $\mu$ is the electrostatic potential, normalised in units of $\phi_0 / 2\pi \tau$; $\eta$ is the friction coefficient and taken to be the (real) dirty-limit value of $\eta = 5.79$ obtained by Schmid [37]. These normalisations lead to the supercurrent $J^s$ normalised in units of $J_0 = B_{c2}/\kappa^2 \mu_0 \xi_s$ where $\mu_0$ is the permeability of free space, and the electric field normalised in units of $J_0 \rho_s$. The three material dependent properties are normalised to those of the superconducting electrode where $\alpha(\mathbf{r})$, the local temperature-dependent condensation energy term; $\beta$, the local self-interaction (nonlinearity) parameter; and $m_i(\mathbf{r})$, the local effective mass tensor of a Cooper pair, which we have here assumed to be diagonal. For simplicity, we shall take $m_i(\mathbf{r})$ and $\alpha(\mathbf{r})$ to be the only spatially varying material dependent parameters and assume the nonlinearity parameter $\beta$ to be constant across the system. The condensation term $\alpha$ can be expressed in terms of the system temperature $T$ and the local critical temperature $T_c(\mathbf{r})$ relative to the critical temperature of the reference superconductor $T_{c,s}$ as

$$\alpha(\mathbf{r}) = \frac{T - T_c(\mathbf{r})}{T - T_{c,s}} \tag{3}$$

such that $\alpha$ is unity in the reference superconductor and negative in normal (non-superconducting) materials. The associated boundary conditions are:

$$(\boldsymbol{\nabla} \times \mathbf{A} - \mathbf{B}_{\mathrm{app}}) \times \hat{\mathbf{n}} = 0 \tag{4}$$

$$(\boldsymbol{\nabla} - \imath \mathbf{A}) \psi \cdot \hat{\mathbf{n}} = -\Gamma_{\mathrm{DG}} \psi \tag{5}$$

where the surface parameter $\Gamma_{\mathrm{DG}}$ is the reciprocal of De Gennes' extrapolation length in units of the coherence length [38] and has the limiting values of 0 for an interface with an insulating surface (or vacuum) and $\pm\infty$ for the interface with a highly conductive surface. In this case, we note that Eqs. (1) and (2) imply continuity of the supercurrent $J_i^s = \frac{1}{m_i(\mathbf{r})} \mathrm{Im}\left[\psi^*(\partial_i - \imath A_i)\psi\right]$. These boundary conditions have previously been used in modelling SNS Josephson junction systems with spatially varying effective mass and critical temperature [39].

However, for many systems of experimental interest that operate in high magnetic fields, Eqs. (1) and (2) are computationally expensive to solve and a further mathematical simplification is needed for 3D simulations. Fortunately, in all high field materials, the (effective) penetration depth is often much larger than all other length scales in the system, and the self field can be neglected relative to the applied magnetic field and current densities, such that the TDGL equations in the high-$\kappa$ limit apply [35]. In this high-$\kappa$ approximation, for an applied magnetic field $B_{\mathrm{app}}$ in the $z$ direction, the normalised magnetic vector potential in the Coulomb gauge ($\boldsymbol{\nabla} \cdot \mathbf{A} = 0$) is expressed as $\mathbf{A} = -B_{\mathrm{app}}(y - w/2)\hat{\mathbf{i}} - \mathbf{K}$, where $\mathbf{K} = K(t)\hat{\mathbf{i}}$ is a spatially invariant parameter required to enforce the Coulomb gauge constraint, and $w$ is the width of the system in the $y$ direction. The governing equations in the high-$\kappa$ approximation, from Eqs. (1) and (2) and the current continuity equation, are

$$\eta\left(\partial_t + \imath \mu\right)\psi = \left[(\boldsymbol{\nabla} - \imath \mathbf{A})^2 + \alpha(\mathbf{r}) - |\psi|^2\right]\psi \tag{6}$$

$$\nabla^2 \mu = \boldsymbol{\nabla} \cdot \mathrm{Im}\left[\psi^*(\boldsymbol{\nabla} - \imath \mathbf{A})\psi\right] \tag{7}$$

$$\partial_t K = J_{\mathrm{app}} - \langle \mathrm{Im}\left[\psi^*(\partial_x - \imath A_x)\psi\right]\rangle \tag{8}$$

where the averaging in Eq. (8) is across the whole domain and at a constant applied magnetic field $B_{\mathrm{app}}$. The gauge constraint $\mathbf{K}$ can be used to determine the average electric field across the domain, since $\partial_t \mathbf{K} = \langle \mathbf{E} \rangle$. The only spatially dependent material parameter in this model is $\alpha(\mathbf{r})$. Spatial variations in the effective mass $m$ are not included, unlike in Eqs. (1) and (2), although a uniform anisotropic $m$ can be accommodated. This formulation is particularly useful for 3D simulations of superconducting systems as the time dependence of the electromagnetic fields is coupled only through the spatially invariant gauge parameter $\boldsymbol{K}$, reducing the computational cost of developing the superconducting state in time [35].

In this work, we also consider SNS junctions analytically and computationally. Following Clem, Eqs. (1) and (2) are considered in terms of gauge-invariant variables: the Cooper pair density $|\psi|^2$, the (super)current density $\boldsymbol{J}^s$ and the gauge-invariant phase $\gamma$ [8], so we can provide physical interpretations of our results. When $m_i(\boldsymbol{r}), \alpha(\boldsymbol{r}), \beta(\boldsymbol{r})$ are only functions of $x$, and solutions for the order parameter are considered in the form $\psi = |\psi|e^{i\theta}$, where $\theta$ is the (non gauge-invariant) phase of the order parameter, the time independent GL equations are [22]

$$\left[\sum_i \left(\partial_i \left[m_i^{-1}(x)\partial_i\right] - m_i^{-1}(x)(\partial_i\gamma)^2\right) + \alpha(\boldsymbol{r}) - \beta(\boldsymbol{r})|\psi|^2\right]|\psi| = 0, \qquad (9)$$

$$\boldsymbol{J}^s = m_i^{-1}(x)|\psi|^2 \boldsymbol{\nabla}\gamma \qquad (10)$$

where

$$\boldsymbol{\nabla}\gamma = \boldsymbol{\nabla}\theta - \boldsymbol{A} \qquad (11)$$

These time-independent solutions are subject to the constraint $\boldsymbol{\nabla} \cdot \boldsymbol{J}^s = 0$. The boundary conditions Eqs. (4) and (5) at an insulating surface become $\widehat{\boldsymbol{n}} \cdot \boldsymbol{J}^s = \widehat{\boldsymbol{n}} \cdot \boldsymbol{\nabla}|\psi| = 0$ [13].

## III. NUMERICAL METHODS FOR SOLVING THE TDGL EQUATIONS FOR JUNCTION SYSTEMS

In this work we use two main simulation codes to solve the TDGL equations for SNS junction systems in simple geometries. For small system sizes in 2D, we will solve the general Eqs. (1) and (2) using our TDGL-2D code, based on the algorithm developed by [40]. For larger systems, and in 3D, we shall solve the simplified TDGL equations in the high $\kappa$ limit, Eqs. (6) to (8), on a GPU using our TDGL-HI$\kappa$ code, an implementation of the 3D TDGL solver developed by [35]. In this section, we shall describe these codes in more detail.

For 2D simulations, we apply the 'link variable' approach used in the explicit method [41] together with the semi-implicit spatial discretisation scheme for the TDGL equations [40] that is generalised to include a spatially dependent effective mass. However although the time evolution of the order parameter $\psi$ is carried out using an adapted version of the Crank-Nicolson algorithm [40], the two components of the magnetic vector potential are then developed in time simultaneously for greater stability when simulating systems with low $\kappa$.

### A. TDGL-2D

#### 1. Spatial Discretisation

Typically, TDGL-2D is used to solve the TDGL equations for systems that are periodic in the direction of current flow in the $x$ direction with periodicity $l$, and bounded in the $y$ direction with a width $w$ such that $y \in [-\frac{w}{2}, \frac{w}{2}]$, at the extremities of which we impose the insulating boundary condition $\Gamma_{\text{DG}} = 0$ using Eq. (5). A schematic of the computational grid, and the relevant dimensions used, are presented in Fig. 1 for the system used to model a typical periodic array of SNS junctions each of thickness $d$. Inside this domain, we specify three regions: a superconducting region of width $w_s$ where $\left(|y| < \frac{w_s}{2}, |x| > \frac{d}{2}\right)$ and in which $\alpha(\boldsymbol{r}) = m_i(\boldsymbol{r}) = 1$; a junction region $\left(|y| < \frac{w_s}{2}, |x| < \frac{d}{2}\right)$ in which $\alpha(\boldsymbol{r}) = \alpha_n$ and $m_i(\boldsymbol{r}) = m_n$; and a coating region $\left(\frac{w_s}{2} < |y| < \frac{w}{2}\right)$ of width $w_{\text{coat}} = (w - w_s)/2$ either side of the junction in which $\alpha(\boldsymbol{r}) = \alpha_{\text{coat}}$ and $m_i(\boldsymbol{r}) = m_{\text{coat}}$. For the 2D simulations presented in this work, $w_{\text{coat}} = 5.0\xi_s$, $\alpha_{\text{coat}} = -10.0$ and $m_{\text{coat}} = 10^8 m_s$ unless otherwise specified.

As the inclusion of a spatially varying effective mass is uncommon in these sorts of simulations, we present the discretisation scheme here explicitly. The simulation space is discretised into a regular grid of nodes at points $\boldsymbol{r}_{i \in [1, n_x], j \in [1, n_y]}$ that are separated by a step size $h_x$ and $h_y$ in the $x$ and $y$ directions. The order parameter $\psi$ is calculated on each node and the 'link variables' $a^x$ and $a^y$ that discretise the magnetic vector potential $\boldsymbol{A}$ are defined on links between nodes,

$$\psi_{i,j} = \psi(\boldsymbol{r}_{i,j}), \qquad a^x_{i,j} = \int_{\boldsymbol{r}_{i,j}}^{\boldsymbol{r}_{i,j}+\hat{i}h_x} A_x \, dx, \qquad a^y_{i,j} = \int_{\boldsymbol{r}_{i,j}}^{\boldsymbol{r}_{i,j}+\hat{j}h_y} A_y \, dy \qquad (12)$$

as shown schematically in the exploded view in Fig. 1. The grid is aligned such that all material boundaries lie between nodes, and thus every node can be identified with a single set of material properties $\alpha, \beta$ and $\eta$. This not true for the effective mass, which is defined on links between nodes,

$$\alpha_{i,j} = \alpha(\boldsymbol{r}_{i,j}), \qquad (m^{-1})^x_{i,j} = h_x^{-1} \int_{\boldsymbol{r}_{i,j}}^{\boldsymbol{r}_{i,j}+\hat{i}h_x} m^{-1}_{xx} \, dx, \qquad (m^{-1})^y_{i,j} = h_y^{-1} \int_{\boldsymbol{r}_{i,j}}^{\boldsymbol{r}_{i,j}+\hat{j}h_y} m^{-1}_{yy} \, dy \qquad (13)$$

on the same grid pattern as the link variables $a^x$ and $a^y$. The observable electric and magnetic fields can be calculated from the link variables,

$$E^\mu_{i,j} = -h_\mu^{-1} \partial_t a^\mu_{i,j} \qquad (14)$$

$$B^z_{i,j} = h_x^{-1} h_y^{-1} \left( a^x_{i,j} - a^x_{i,j+1} - a^y_{i,j} + a^y_{i+1,j} \right) \qquad (15)$$

as required. With these definitions, the spatial discretisation of Eqs. (1) and (2) in the zero electric potential gauge ($\mu = 0$) accurate to second order is

$$\eta \partial_t \psi_{i,j} = h_x^{-2} \left[ (m^{-1})^x_{i-1,j} e^{ia^x_{i-1,j}} \psi_{i-1,j} - \left( (m^{-1})^x_{i-1,j} + (m^{-1})^x_{i,j} \right) \psi_{i,j} + (m^{-1})^x_{i,j} e^{-ia^x_{i,j}} \psi_{i+1,j} \right]$$
$$+ h_y^{-2} \left[ (m^{-1})^y_{i,j-1} e^{ia^y_{i,j-1}} \psi_{i,j-1} - \left( (m^{-1})^y_{i,j-1} + (m^{-1})^y_{i,j} \right) \psi_{i,j} + (m^{-1})^y_{i,j} e^{-ia^y_{i,j}} \psi_{i,j+1} \right]$$
$$+ \left( \alpha_{i,j} - |\psi_{i,j}|^2 \right) \psi_{i,j} \qquad (16)$$

$$\partial_t a^x_{i,j} = \kappa^2 m^x_{i,j} h_y^{-2} \left( a^x_{i,j+1} - 2a^x_{i,j} + a^x_{i,j-1} - a^y_{i+1,j} + a^y_{i,j} + a^y_{i+1,j-1} - a^y_{i,j-1} \right) + \text{Im} \left[ \psi^*_{i,j} e^{-ia^x_{i,j}} \psi_{i+1,j} \right] \qquad (17)$$

$$\partial_t a^y_{i,j} = \kappa^2 m^y_{i,j} h_x^{-2} \left( a^y_{i+1,j} - 2a^y_{i,j} + a^y_{i-1,j} - a^x_{i,j+1} + a^x_{i,j} + a^x_{i-1,j+1} - a^x_{i-1,j} \right) + \text{Im} \left[ \psi^*_{i,j} e^{-ia^y_{i,j}} \psi_{i,j+1} \right] \qquad (18)$$

where $m^x_{i,j} = 1/(m^{-1})^x_{i,j}$ and $m^y_{i,j} = 1/(m^{-1})^y_{i,j}$. We note that these spatially discretised equations remain gauge invariant [40].

Imposing periodic boundary conditions in the $x$ direction requires $\psi_{0,j} \equiv \psi_{n_x,j}$, $\psi_{n_x+1,j} \equiv \psi_{1,j}$, $a^x_{0,j} \equiv a^x_{n_x,j}$, $a^x_{n_x+1,j} \equiv a^x_{1,j}$, $a^y_{0,j} \equiv a^y_{n_x,j}$, $a^y_{n_x+1,j} \equiv a^y_{1,j}$. In the $y$ direction, the boundary conditions Eqs. (4) and (5) are implemented using a ghost point method [42]:

$$\eta \partial_t \psi_{i,1} = h_x^{-2} \Big[ (m^{-1})^x_{i-1,1} e^{ia^x_{i-1,1}} \psi_{i-1,1} + (m^{-1})^x_{i,1} e^{-ia^x_{i,1}} \psi_{i+1,1}$$
$$- \left( (m^{-1})^x_{i-1,1} + (m^{-1})^x_{i,1} \right) \psi_{i,1} \Big]$$
$$+ h_y^{-2} \Big[ \left( h_y \Gamma_{\text{DG}} - (m^{-1})^y_{i,1} \right) \psi_{i,1} + (m^{-1})^y_{i,1} e^{-ia^y_{i,1}} \psi_{i,2} \Big]$$
$$+ \left( \alpha_{i,1} - |\psi_{i,1}|^2 \right) \psi_{i,1} \tag{19}$$

$$\eta \partial_t \psi_{i,n_y} = h_x^{-2} \Big[ (m^{-1})^x_{i-1,n_y} e^{ia^x_{i-1,n_y}} \psi_{i-1,n_y} + (m^{-1})^x_{i,n_y} e^{-ia^x_{i,n_y}} \psi_{i+1,n_y}$$
$$- \left( (m^{-1})^x_{i-1,n_y} + (m^{-1})^x_{i,n_y} \right) \psi_{i,n_y} \Big]$$
$$+ h_y^{-2} \Big[ (m^{-1})^y_{i,n_y-1} e^{ia^y_{i,n_y-1}} \psi_{i,n_y-1} + \left( h_y \Gamma_{\text{DG}} - (m^{-1})^y_{i,n_y-1} \right) \psi_{i,n_y} \Big]$$
$$+ \left( \alpha_{i,n_y} - |\psi_{i,n_y}|^2 \right) \psi_{i,n_y} \tag{20}$$

$$\partial_t a^x_{i,n_y} = \kappa^2 m^x_{i,n_y} h_y^{-2} \left( -a^x_{i,n_y} + a^x_{i,n_y-1} + a^y_{i,n_y-1} - a^y_{i+1,n_y-1} - h_x h_y (B_{\text{app}} + \frac{w}{2}\mu_0 J_{\text{app}}) \right)$$
$$+ \text{Im}\left[ \psi^*_{i,n_y} e^{-ia^x_{i,n_y}} \psi_{i+1,n_y} \right] \tag{21}$$

$$\partial_t a^x_{i,1} = \kappa^2 m^x_{i,1} h_y^{-2} \left( -a^x_{i,1} + a^x_{i,2} + a^y_{i,1} - a^y_{i+1,1} + h_x h_y B_{\text{app}} - \frac{w}{2}\mu_0 J_{\text{app}} \right)$$
$$+ \text{Im}\left[ \psi^*_{i,1} e^{-ia^x_{i,1}} \psi_{i+1,1} \right] \tag{22}$$

For convenience, we will define the multi-indices $\mu$ and $\nu$ that specify the link variable $a^\mu = a^u_{i,j}$ and $\psi^\nu = \psi_{i,j}$ respectively as

$$\mu(i,j,u) = \begin{cases} j + (i-1)n_y & \text{for } u = x \\ n_x n_y + i + (j-1)n_x & \text{for } u = y \end{cases}, \quad \nu(i,j) = i + (j-1)n_x \tag{23}$$

The above equations can be simplified into (using the Einstein summation convention [43]):

$$\partial_t a^\mu = J^\mu_{\mu'} a^{\mu'} + S(\{a,\psi\})^\mu \tag{24}$$

$$\partial_t \psi^\nu = L^\nu_{\nu'}(\{a\}) \psi^{\nu'} + N(\{\psi\})^\nu \tag{25}$$

where the nonlinear terms

$$N(\{\psi\})^\nu = (\eta^{-1})^\nu \left( \alpha^\nu - |\psi^\nu|^2 \right) \psi^\nu \tag{26}$$

and

$$S(\{a,\psi\})^\mu = \begin{cases} \text{Im}\left[ \psi^*_{i,j} e^{-ia^\mu} \psi_{i+1,j} \right] & \text{for } u(\mu) = x \\ \text{Im}\left[ \psi^*_{i,j} e^{-ia^\mu} \psi_{i,j+1} \right] & \text{for } u(\mu) = y \end{cases} \tag{27}$$

## 2. Temporal Discretisation

For evolving $\{a, \psi\}$ the adapted Crank-Nicolson algorithm [40] is known to be unconditionally stable for purely linear sets of equations [42], although stability is not guaranteed in the nonlinear case. Unlike the explicit scheme of Gropp et al. [41], that uses the computational variables $\{U\} = \{\exp(-\imath a)\}$ instead of $\{a\}$ directly, numerical errors of schemes based on [40] will increase for long simulations of periodic systems in resistive states, as the magnitude of $\{a\}$ can grow large over time as a result of Eq. (14), and slow or even prevent convergence. However, as we are predominately interested in the critical current density $J_c$ and the onset of persistent resistive states in the system, this does not significantly limit the simulations presented here, and this consideration is outweighed by the reduction in simulation time possible using the longer timesteps that Crank-Nicolson approach permits as a result of its greater stability properties.

Applying a Crank-Nicolson approach on all terms in Eqs. (16) and (17) we can relate the computational variables at timestep $n+1$ and $n$ by:

$$\frac{a_{n+1}^\mu - a_n^\mu}{\delta t} = \frac{1}{2}\left[J_{\mu',n}^\mu a_n^{\mu'} + J_{\mu',n+1}^\mu a_{n+1}^{\mu'} + S(\{a_n, \psi_n\})^\mu + S(\{a_{n+1}, \psi_{n+1}\})^\mu\right] \tag{28}$$

$$\frac{\psi_{n+1}^\nu - \psi_n^\nu}{\delta t} = \frac{1}{2}\left[L_{\nu'}^\nu(\{a_n\})\psi_n^{\nu'} + L_{\nu'}^\nu(\{a_{n+1}\})\psi_{n+1}^{\nu'} + N(\{\psi_n\})^\nu + N(\{\psi_{n+1}\})^\nu\right] \tag{29}$$

Rearranging, we arrive at a pair of coupled, nonlinear equations to be solved for our unknown variables $\{a_{n+1}, \psi_{n+1}\}$ at each timestep,

$$\mathcal{J}_{\mu',n+1}^{\mu,-} a_{n+1}^\mu = \mathcal{J}_{\mu',n}^{\mu,+} a_n^{\mu'} + \frac{\delta t}{2}\left[S(\{a_n, \psi_n\})^\mu + S(\{a_{n+1}, \psi_{n+1}\})^\mu\right] \tag{30}$$

$$\mathcal{L}_{\nu'}^{\nu,-}(\{a_{n+1}\})\psi_{n+1}^\nu = \mathcal{L}_{\nu'}^{\nu,+}(\{a_n\})\psi_n^{\nu'} + \frac{\delta t}{2}\left[N(\{\psi_n\})^\nu + N(\{\psi_{n+1}\})^\nu\right] \tag{31}$$

where we have defined

$$\mathcal{J}_{\mu',n}^{\mu,\pm} = \delta_{\mu'}^\mu \pm \frac{\delta t}{2} J_{\mu',n}^\mu, \qquad \mathcal{L}_{\nu'}^{\nu,\pm}(\{a_n\}) = \delta_{\nu'}^\nu \pm \frac{\delta t}{2} L_{\nu'}^\nu(\{a_n\}) \tag{32}$$

However, as these equations are nonlinear, an iterative method must be employed at each timestep. Fortunately, since the timescale for the evolution of $\{a\}$ is much shorter than $\{\psi\}$ since usually $\kappa^2 \gg \eta^{-1}$, we have applied a block Gauss-Seidel approach to the fully coupled system [42]. Denoting the $m^{\text{th}}$ iteration of our set of unknowns by $\{a_{n+1}^{(m)}, \psi_{n+1}^{(m)}\}$ we have

$$\mathcal{L}_{\nu'}^{\nu,-}(\{a_{n+1}^{(m)}\})\psi_{n+1}^\nu = \mathcal{L}_{\nu'}^{\nu,+}(\{a_n\})\psi_n^{\nu'} + \frac{\delta t}{2}\left[N(\{\psi_n\})^\nu + N(\{\psi_{n+1}^{(m)}\})^\nu\right], \tag{33}$$

$$\mathcal{J}_{\mu',n+1}^{\mu,-} a_{n+1}^{\mu,(m+1)} = \mathcal{J}_{\mu',n}^{\mu,+} a_n^{\mu'} + \frac{\delta t}{2}\left[S(\{a_n, \psi_n\})^\mu + S(\{a_{n+1}^{(m)}, \psi_{n+1}^{(m+1)}\})^\mu\right], \tag{34}$$

where we set $\{a_{n+1}^{(0)}, \psi_{n+1}^{(0)}\} = \{a_n, \psi_n\}$. However, unlike [40], we do not separate Eq. (34) into two iteration steps, as the timescale for the evolution of $\{a^x\}$ and $\{a^y\}$ are similar magnitudes, which can lead to oscillatory behaviour of the iteration scheme with a block Gauss-Seidel approach and unreliability of convergence [42], Equation (33) is solved directly and more quickly in two steps using the method of fractional steps to decompose the linear operator $\mathcal{L}_{\nu'}^{\nu,-}(\{a_{n+1}^{(m)}\})$ into the product $\mathcal{L}_{X,\nu'}^{\nu,-}(\{a_{n+1}^{(m)}\})\mathcal{L}_{Y,\nu'}^{\nu,-}(\{a_{n+1}^{(m)}\})$ of two simpler operators containing difference terms in one dimension only. In this geometry, $\mathcal{L}_{X,\nu'}^{\nu,-}$ is a cyclic tridiagonal matrix and $\mathcal{L}_{Y,\nu'}^{\nu,-}$ is a banded tridiagonal matrix, for which fast solution methods are available. Cyclic tridiagonal systems are solved using a Sherman-Morrisson algorithm [44] with the tridiagonal solver provided by the LAPACK package. Equation (34) is solved in one solution step using the Intel MKL PARDISO direct parallel sparse solver. Factorisation and analysis of the operator $\mathcal{J}_{\mu',n+1}^{\mu,-}$ need only be performed once as the values are time-independent, with the exception of boundary terms that can be grouped with nonlinear terms in Eq. (34).

Convergence was achieved by solving Eqs. (33) and (34) alternately until the maximum residual $\varepsilon$, defined by

$$\varepsilon = \max\left\{\left|a_{n+1}^{(m+1)} - a_{n+1}^{(m)}\right|, \left|\text{Re}\left[\psi_{n+1}^{(m+1)} - \psi_{n+1}^{(m)}\right]\right|, \left|\text{Im}\left[\psi_{n+1}^{(m+1)} - \psi_{n+1}^{(m)}\right]\right|\right\}, \quad (35)$$

satisfied $\varepsilon < 10^{-7}$ at each time step.

### 3. Critical Current Simulations

In order to extract values for the critical current density $J_c$ of the system in a given applied field $B_{\text{app}}$, we successfully established a procedure to determine the lowest applied current density $J_{\text{app}}$ at which a persistent resistive state is observed. We followed the experimental approach [45] and used an arbitrary electric field criterion $E_c$ written interms of $E_D$, which corresponds to the average electric field in the system when the superconductor is normal and carrying the zero-field Ginzburg–Landau depairing current density $J_D$, such that

$$E_D = \kappa^2 \rho_{\text{av}}^x J_D, \quad (36)$$

where

$$\rho_{\text{av}}^x = \frac{w}{w_s} \frac{1}{n_x} \sum_{i=1}^{n_x} \frac{n_y}{\sum_{j=1}^{n_y}\left[(m^{-1})_{i,j}^x\right]}, \qquad J_D = \frac{2}{3\sqrt{3}} J_0 \quad (37)$$

where $\rho_{\text{av}}^x$ represents the average resistivity of the system in the $x$-direction, normalised to the resistivity of a system in the $x$-direction containing only the superconductor in its normal state. As the critical current density of the superconductor can be highly hysteretic, the system was always first initialised in the Meissner state throughout ($\psi = 1$, $\boldsymbol{A} = 0$) for all simulations. The external magnetic field $B$ $(y = \pm\frac{w}{2})$ was then increased at a rate of $5 \times 10^{-2} B_{c2} \tau^{-1}$ up to the desired value $B_{\text{app}}$. Following this magnetic field ramp, for our 2D (3D) simulations the applied current density $J_{\text{app}}$ was increased (decreased) in a series of logarithmically spaced steps, starting from $10^{-6} J_D$. If the average electric field in the system exceeded the electric field criterion, typically $E_c = 10^{-5} E_D$, the applied current was held constant. When the average electric field continued to persist above $E_c$ for longer than the hold time $t_{\text{hold}}$, typically taken as $5 \times 10^4 \tau$, the system was determined to have entered a persistent resistive state and $J_{\text{app}}$ at this point is taken to be the critical current density of the system.

An example of the time evolution of the applied current density and average electric field used to extract $J_c$ from the simulation is displayed in Fig. 2. The rapid jumps in the average electric field in the system $\langle E_x \rangle$ below the critical current $(t < 1.1 \times 10^4)$ are associated with the imposed current steps and the associated steps in the rate of change of the magnetic field in the system. To make the generation of a full $J_c(B_{\text{app}})$ characteristic more efficient, we also simulate $J_c$ at different applied fields in parallel, since the simulations for the critical current at given applied fields are independent of one another.

## B. TDGL-HI$\kappa$

### 1. GPU Algorithm

Unfortunately, for large grid sizes, solutions of Eq. (34) with the link variables $\{a_{x,y}\}$ being updated in a single step become prohibitively expensive, and thus the algorithm described in Section III A scales poorly for 3D systems. In this limit, Sadovskyy et al. have developed a scalable GPU accelerated algorithm to solve Eqs. (6) to (8) to investigate the effect of pinning structures in 3D superconducting systems [35]. For 3D simulations, we have written and implemented a TDGL solver (TDGL-HI$\kappa$) using the algorithm in [35], to investigate $J_c$ in large scale polycrystalline systems.

The order parameter $\psi$, the electrostatic potential $\mu$, and the gauge parameter $K$ are updated successively at each timestep, with $\psi$ and $\mu$ solved for iteratively as described in [35] until $|\psi_{n+1} - \psi_n|^2 < 10^{-5}$ and $\left|\nabla^2 \mu - \nabla \cdot \text{Im}\left[\psi^*(\nabla - \imath \boldsymbol{A})\psi\right]\right|^2 < 10^{-5}$ at every mesh point. $K$ is integrated forward in time using a second order Runge-Kutta algorithm [44]. Local order parameter fluctuations may also be included for investigations of vortex creep by adding a temperature dependent noise term $\zeta = \zeta_1 + \imath \zeta_2$ to the right hand side of Eq. (6) [35]. $\zeta_1$ and $\zeta_2$ are independent random variables at each timestep, taken from the uniform distribution in the interval between $\zeta_{\max} = \left(3\eta T_f h_t h_x h_y h_z / \xi_s^3 \tau\right)^{1/2}$ In this work however, we set $T_f = 10^{-6}$, which is sufficiently small so as to minimise creep effects that may complicate the determination of $J_c$ and corresponds to nearly zero thermal noise for vortex flow [46], but sufficiently large to speed up relaxation of the order parameter when the system is out of equilibrium, such as immediately after initialisation. Insulating or (quasi)periodic boundary conditions can be applied at the edges of the simulation domain in any (or all) spatial dimensions [35]. For a periodic domain of size $L_x, L_y, L_z$ in the $x, y$ and $z$ dimensions respectively with a magnetic field applied along the $z$ axis, periodic boundary conditions can be applied to $\psi$ at the edges of the domain in the $x$ and $z$ dimensions, and quasiperiodic boundary conditions (QBC) on $\psi$ in the $y$ dimension, as described in [35]. We found QBC particularly useful in 3D to eliminate surface effects from masking bulk critical currents in computationally accessible system sizes.

### 2. Critical Current Simulations

For 3D simulations, we follow the $J_c$ determination method employed in [47], and ramp the applied current down in steps from the resistive to the superconducting state. At each current step, the current is held for $t_{\text{hold}}$ and the spatially averaged electric field in the superconductor $E_x$ is averaged over the second half of the hold step, after transient effects from stepping the current have decayed away. Typically $t_{\text{hold}} = 10.0 \ \tau$. The critical current density $J_c$ is then taken to be the highest current at which the time-averaged and spatially-averaged $E_x$ is less than the electric field criterion $E_c = 10^{-5} \rho J_0$.

## IV. REVIEW OF WEAKLY COUPLED SNS JUNCTIONS IN LOW MAGNETIC FIELDS ($\alpha_n d \gg \xi_s$)

We shall now review analytic expressions for the critical current density across weakly coupled Josephson junction systems in low fields from the literature. For simplicity, we will restrict this analytic discussion to solving Eqs. (9) and (10) in 2D, valid for thin superconducting films, or volumes of superconducting system in which all vortices are parallel to one another and the junction plane. The system geometry is shown in Fig. 1.

It is convenient to classify junctions depending on the relative size of the junction dimensions compared to the superconductor coherence length $\xi_s$ and penetration depth $\lambda_s$. We shall categorise junctions by their width $w_s$ transverse to the direction of current flow relative to length scales of the superconductor, into 'very narrow' junctions, with $w_s \ll \xi_s, \lambda_s$; 'narrow' junctions, with $\xi_s \ll w_s \ll \lambda_s$, and 'wide' junctions, with $w_s \gg \xi_s, \lambda_s$. In addition, it shall be useful to further subcategorize these systems as 'thin' or 'thick' junctions between weakly coupled ($-\alpha_n d \gg 1$) superconductors depending on whether its thickness $d$ in the direction of current flow is much smaller or much larger than the superconducting coherence length $\xi_s$. Here we consider these different types of junctions in turn.

### A. Very Narrow Junctions $w \ll \xi_s$

For very narrow junctions with insulating boundary conditions, $w = w_s \ll \xi_s$, such that the boundary condition $\hat{\boldsymbol{n}} \cdot \boldsymbol{\nabla}|\psi| = 0$ implies $[\partial_y |\psi|]_{-w/2}^{w/2} = 0$. In this very narrow junctions case, no vortices are stable inside the structure and the magnitude of the order parameter $|\psi|$ is approximately constant along the $y$ direction. Hence, we can integrate Eq. (9) over the junction width in the $y$ direction, apply the mean value theorem, and replace $\psi$ with its average in the $y$ direction $f = \frac{1}{w} \int_{-w/2}^{w/2} |\psi| \, dy$ and the components of $\boldsymbol{J}^s$ by their equivalent average $\langle j_i^s \rangle_y = \frac{1}{w} \int_{-w/2}^{w/2} (J_i^s) \, dy$. In the limit where the applied magnetic field is much less than the self field, $J_y^s = 0$ from the insulating boundary conditions, and $J_x^s$ is independent of $y$. Eq. (9) is then reduced to an equation in only one variable $x$. Using Eq. (10) gives

$$\partial_x \left( m^{-1}(x) \partial_x f \right) + \left[ \alpha(x) - \beta(x) f^2 - \frac{m(x) \langle j_x^s \rangle_y^2}{f^4} \right] f = 0 \tag{38}$$

#### 1. Thin Junctions $d \ll \xi_s$

The critical current in the thin junction limit, where $d \ll \xi_s$, has been solved in weakly coupled limit by [48] and investigated numerically in the strongly coupled case by [49]. To keep things simple here, we take $\beta(x)$ and $m^{-1}(x)$ as constant across the system. Equation (38) is written as

$$\partial_x^2 f + \left[ 1 - (1 - \alpha(x)) - f^2 - \frac{\langle j_x^s \rangle_y^2}{f^4} \right] f = 0 \tag{39}$$

Since $f$ and $\langle j_x^s \rangle_y$ are continuous across the S/N interface in this case, a constraint between $\partial_x f$ and $f$ at the interface in the limit where $d \ll \xi_s$ can easily be found, by integrating Eq. (39) in the $x$ direction across the normal region. Assuming $f$ is symmetric across the junction and $1 - \alpha_n \sim O(d^{-1})$ or larger, then to leading order in $d/\xi_s$

$$2 f'_{d/2} = d(1 - \alpha_n) f_{d/2}, \tag{40}$$

where $f_{d/2} = f(x = d/2)$ and $f'_{d/2} = \partial_x f(x = d/2)$.

As shown by [48], in the weak coupling case, when the critical current density of the junction is much less than the critical current density of the bulk superconductors, $\lim_{x \to \infty} \{f\} = 1$ and $\lim_{x \to \infty} \{f'\} = 0$, so that integrating Eq. (39) from the S/N interface to a point far from the junction yields

$$f'^2_{d/2} + f^2_{d/2} - \frac{f^4_{d/2}}{2} + \frac{\langle j_x^s \rangle_y^2}{f^2_{d/2}} = \frac{1}{2} \tag{41}$$

Substituting $f'_{d/2}$ from Eq. (40) into Eq. (41) and neglecting the highest order terms in the small parameter $V_0^{-1} = 1/d(1-\alpha_n)$ gives

$$f_{d/2}^2 = V_0^{-2} + V_0^{-1}\sqrt{V_0^{-2} - 4\langle j_x^s\rangle_y^2} \tag{42}$$

From the discriminant, we see that for $f$ to remain positive and real at the S/N interface, $\langle j_x^s\rangle_y \leq V_0/2$. This gives us the condition for the maximum critical current that can flow through the junction that we denote as $J_{\text{DJ}}$ where,

$$\lim_{d\ll\xi_s}\{J_{\text{DJ}}(B_{\text{app}}=0)\} = J_0\frac{\xi_s}{2d(1-\alpha_n)} \tag{43}$$

$J_{\text{DJ}}$ is the equivalent of the Ginzburg–Landau depairing current density for a junction system; the maximum lossless critical current density that can flow, above which superconductivity in the system is destroyed.

2. *Thick Junctions $d \gg \xi_s$*

For thick junctions, the critical current density for thick junctions has been solved by [14] from Eq. (38). In the superconductor regions, with $f_s = f$, $j_x = \langle j_x^s\rangle_y$, Eq. (38) can be written

$$\partial_x^2 f_s + \left[1 - f_s^2 - \frac{j_x^2}{f_s^4}\right]f_s = 0, \tag{44}$$

whereas inside the normal region, Eq. (38) can be rescaled with $u = x\sqrt{-\alpha_n m_n/m_s}$, $f_n = -f\sqrt{-\beta_n/\alpha_n}$ and $j_u = \langle j_x^s\rangle_y \beta_n\sqrt{m_n/m_s}(-\alpha_n)^{-3/2}$ to give

$$-\partial_u^2 f_n + \left[1 - f_n^2 + \frac{j_u^2}{f_n^4}\right]f_n = 0. \tag{45}$$

Equations (44) and (45) can be solved analytically for the magnitude of the order parameter in terms of Jacobi elliptic functions. For thick junctions, where the order parameter at the centre of the junction is much smaller than that at the S/N boundary, the maximum critical current density for this system of equations can be obtained in the form [14]:

$$\lim_{d\gg\xi_s}\{J_{\text{DJ}}(B_{\text{app}}=0)\} = 4J_0\frac{1-\sqrt{1-sf_{d/2}^2}}{sv}\exp\left(-\frac{d}{\xi_n}\right), \qquad f_{d/2}^2 = \frac{v^2+1-\sqrt{v^2(2-s)+1}}{v^2+s} \tag{46}$$

where

$$v = \frac{m_n\xi_n}{m_s\xi_s}, \qquad s = -\frac{\beta_n}{|\alpha_n|}, \qquad \xi_n = \sqrt{\frac{m_s}{m_n}\frac{1}{|\alpha_n|}}\xi_s \tag{47}$$

Here we have included the nonlinearity parameter inside the junction $\beta = \beta_n$ in the normalisation for generality (but in this work taken $\beta_n = 1$) which implies from Eq. (47) that $s < 0$, in contrast to the numerical solutions studied by [14]. We note that in the linearised limit ($s \to 0$) this zero-field critical current reduces to the limit found by [50]. Furthermore, in the limit $v^2 \to -s$, then $f_{d/2}^2 \to 1/2(1-\alpha_n)$ and for the specific case $f_{d/2}^2 \to 0$ we find Eq. (46) reduces to the well-known form

$$J_{\text{DJ}} = J_0\frac{\xi_n}{\xi_s}\exp\left(-\frac{d}{\xi_n}\right) \tag{48}$$

first found by De Gennes for SNS junctions [12] to first order and by Jacobson [13] through a similar approach.

### B. Narrow Junctions, $\lambda_s \gg w_s \gg \xi_s$

For narrow junctions, vortices penetrate the junction even in low fields. Consideration of low field solutions to the Ginzburg-Landau equations of the form $\psi = |\psi|e^{i\theta}$ led Josephson to propose his relation:

$$J = J_{\mathrm{DJ}} \sin(\Delta\gamma) \tag{49}$$

where $J$ is the average current density along a contour between two points across the junction, $J_{\mathrm{DJ}}$ is a constant and $\Delta\gamma$ is the difference in the gauge invariant phase between the points. The general solutions for the critical current density derived from the Ginzburg Landau equations have been compared to those generated using the Josephson relation in low magnetic fields [49]. The critical current density from Eq. (43) approximates the general solution well in the weak coupling limit $V_0 > 8$, but breaks down when $V_0 \to 0$ and $J_{\mathrm{DJ}} \to J_{\mathrm{D}}$ [49].

Low field solutions for the gauge invariant phase difference $\Delta\gamma(y)$ in thin films have been found by Clem [8]. Whilst the original formalism was developed for thin films, it remains applicable the narrow 2D systems considered here since in both cases, $\psi$ is independent of $z$ and the local magnetic field can be taken to be equal to the applied field as $w < \lambda_s$. These low field solutions for the gauge invariant phase difference $\Delta\gamma(y)$ and average critical current density across a narrow junction [8] are given by

$$\Delta\gamma(y) = \Delta\gamma(0) + B_{\mathrm{app}} y d_{\mathrm{eff}} + \frac{8 B_{\mathrm{app}}}{w_s} \sum_{n=0}^{\infty} \frac{(-1)^n}{k_n^3} \tanh(k_n l_s/2) \sin(k_n y), \qquad k_n = (2n+1)\pi/w_s \tag{50}$$

$$J_c = \max_{\varphi(0)} \left\{ \frac{1}{w_s} \left| \int_{w_s/2}^{w_s/2} dy \, [J_{\mathrm{DJ}}(0) \sin(\Delta\gamma(y))] \right| \right\} \tag{51}$$

where $J_{\mathrm{DJ}}(0)$ is the current density in zero field. In this case, $\gamma(0) = \pm\pi/2$ when the current through the junction is maximised for all ratios of $l_s/w_s$ [8]. In order to improve agreement between our computation and Eq. (50), we have included a term for the effective junction thickness $d_{\mathrm{eff}}$ (which we find below to be $d_{\mathrm{eff}} \approx 2\xi_s$ in the weak coupling limit). This term accounts for the finite size of the junction and the reduction in the order parameter on a length scale of order $\xi_s$ close to the junction. This addition better describes thin junctions (i.e the limit considered by [8]). For consistency, we define the effective length of the S regions in the direction of current flow to be $l_s = l - d_{\mathrm{eff}}$.

Equation (50) demonstrates that the screening currents close to the edges of the junction depend sensitively on the aspect ratio of the S regions and determine the magnetic field dependence of $J_c$. For short electrodes, when $l_s \ll w_s$, current flow close to the N region is mostly parallel to the junction across the whole length, Josephson vortices in the junction are spaced approximately equally along the junction width, and the critical current density has a sinc-like functional form:

$$J_c = J_{\mathrm{DJ}}(0) \frac{\phi_0}{\pi \phi_{\mathrm{short}}} \left| \sin\left(\frac{\pi \phi_{\mathrm{short}}}{\phi_0}\right) \right|, \qquad \phi_{\mathrm{short}} = w_s l_s B_{\mathrm{app}}. \tag{52}$$

where $J_{\mathrm{DJ}}(0)$ is the current density in zero field. In contrast, for long electrodes, $l_s \gg w_s$, and $d_{\mathrm{eff}} \to 0$, screening currents flowing in the superconductors curve away from the junction across most of the junction width. As a result, Josephson vortices close to the edges are spaced further apart at the edges than at the centre, and larger current densities can be carried in the edge regions. In this case, the critical current density can be approximated by a Bessel-like functional form where

$$J_c = J_{DJ}(0)\left|\mathcal{J}_0\left(\frac{\pi\phi_{\text{long}}}{\phi_0}\right)\right|, \qquad \phi_{\text{long}} = 14\zeta(3)B_{\text{app}}w_s^2/\pi^3. \tag{53}$$

where $\mathcal{J}_0$ is the Bessel function of the first kind of order 0, and $\zeta(3) = 1.202$. Eq. (53) shows the distance between the cores of the vortices in the junction, $a_J$, is given by $a_J \approx 1.84\phi_0/B_{\text{app}}w_s$. To identify the fraction of the width contributing to the *net* critical current, we note that the maxima of Eq. (51), $J_c^{\text{peak}}$, can be approximated when $w_s \approx l_s$ using:

$$J_c^{\text{peak}} \approx c_0 \left(\frac{\phi_0}{Bw_s^2}\right)^{c_1} J_{DJ}(0) \tag{54}$$

We find empirically that over a large range of aspect ratios, the field dependence of $J_c^{\text{peak}}$ most closely follows the Bessel function field dependence where for example when $w_s \approx l_s$, $c_0 \approx c_1 \approx 0.6$ and over a range of aspect ratios for the electrodes, $c_0 \approx 0.35/c_1$ is quite robust. As noted by [10, 21], the reduction of the critical current with applied field when many vortices are present in the junction is slower when $w_s \ll l_s$ compared to when $l_s \ll w_s$, since the asymptotic behaviour of the Bessel-like function (Eq. (53)) has $J_c \sim B_{\text{app}}^{-1/2}$ whereas the sinc-like function has $J_c \sim B_{\text{app}}^{-1}$ (Eq. (52)). A comparison between the critical current density determined from Eqs. (50) and (51) and the critical current density obtained from our 2D TDGL simulations is shown in Fig. 3 for a system with $w_s \gg l_s$ (upper panel) and $w_s \ll l_s$ (lower panel). In both cases, we take $d_{\text{eff}} \approx 2\xi_s$. The 2D TDGL simulations $J_c$ from both TDGL-2D and TDGL-HI$\kappa$ show excellent agreement with each other and the analytic expressions derived from Eqs. (50) and (51) in low fields. At these applied fields, no vortices exist in the S regions, and current flow is laminar within them. In the lower panel of Fig. 3, simulations of $J_c$ obtained from TDGL-2D for larger system widths at $B = 0.2B_{c2}$ still follow the prediction of Eqs. (50) and (51), but with larger scatter as a consequence of vortices in the S regions that distort the interference pattern of the computed system from the analytic prediction [21].

For completeness, we checked our results against a smaller grid step size $0.1\xi_s$ and confirmed little change in $J_c(B)$ values. Throughout this work, a standard grid step size of $0.5\xi_s$ was chosen since it gave the optimal trade-off between accuracy and computation time. We also checked the sensitivity of the results in this section to having a highly resistive coating, rather than an insulator, at the edges of the junction system. This coating allows the order parameter at the superconductor/coating interface to decay into the coating region which affects the critical current characteristics in-field. The simulation data shown in Fig. 4 show that insulating surface conditions are found if the effective mass in the coating material is greater than around 30 times the maximum effective mass in the rest of the system.

### C. Wide Junctions, $w_s \gg \lambda_s, \xi_s$

For completeness, we briefly consider wide junctions. In wide junctions between weakly coupled superconductors in low fields, with $w_s \gg \lambda_s$ but still smaller than the Josephson penetration depth $\lambda_J$, the screening currents that flow around the S regions screen most of the applied magnetic field. In low fields, when vortices penetrate the N region but the S regions are in the Meissner state in the bulk, away from the junction edges $\varphi(y) = Byd_{\text{eff}}/B_{c2}\xi_s^2$ along the junction which now has an effective thickness $d_{\text{eff}} = (2\lambda + d)$, and the net critical current density of the junction is once again given by the sinc-like pattern described by Eq. (52) but with $\phi = Bw_s(2\lambda_s + d)/B_{c2}\xi_s^2$ [16]. This crossover has been investigated analytically by [18]. For the very widest junctions, larger than the Josephson penetration depth $\lambda_J$, the current flowing across the junction itself can be large enough to screen the applied magnetic field, reducing the local magnetic field at the centre of the junction. In general, the field dependence of the critical current is multivalued and requires the solution of trancendental equations [15], with solution branches depending on the number of whole vortices in the junction [19]. Physically, this arises because states with different numbers of vortices and different critical current density can be stable at the same applied field due to the presence of a surface barrier in the system, and, in general, these states have different critical currents. However, in the high field limit, when the applied field is sufficiently large such that there are many vortices in the junction, the solutions tend towards the sinc-like pattern of Eq. (52) with $\phi = Bw_s(2\xi_s + d)/B_{c2}\xi_s^2$ when $w_s, l_s \gg \lambda_J$ and the high-field envelope of the critical current density once again follows $J_c \sim B_{\text{app}}^{-1}$ [15, 50].

## V. JOSEPHSON JUNCTIONS IN HIGH MAGNETIC FIELDS

In this section, we derive new analytic expressions for the critical current density of very narrow Josephson junctions ($w < \xi_s$), that are valid across the entire range of applied magnetic fields, up to the upper critical field of the system. We shall then use these expressions to form approximations for the dependence of the critical current density on applied magnetic fields for very narrow and narrow junction systems ($\xi_s \ll w \ll \lambda_s$). First we consider current flow within the junction from screening currents and from the injected currents. Integrating around a thin closed rectangular loop inside the system using Eq. (11) with the lower path along the $x$-axis and the upper path at $y$ gives

$$\oint \nabla \gamma \cdot dl = \oint \nabla \theta \cdot dl - \oint B \cdot dS \tag{55}$$

after applying Stoke's theorem on the magnetic vector potential term. For any choice of gauge, the first closed integral on the RHS in $\theta$ is $2\pi n$ where $n$ is the number of vortex cores inside the closed contour, from the requirement that the order parameter magnitude be a single valued function. We assume that the order parameter magnitude is symmetric about both the $y$-axis and $x$-axis, that the screening currents and hence $\partial_y \gamma$ are both antisymmetric about these axes, and to first order the transport current is uniform along the $y$-axis, such that $\langle j_x^s \rangle_y = m_x^{-1}(x) f^2 \partial_x \gamma (y = 0)$ from Eq. (10). Given no vortex cores exist in the narrow system ($n = 0$), and taking the sections of the contour in Eq. (55) that are parallel to the $x$ axis to be sufficiently short relative to the coherence length $\xi$ leads to the gauge invariant result

$$\partial_x \gamma(y) - \frac{\langle j_x^s \rangle_y}{f^2 m_x^{-1}(x)} = \frac{B_{\text{app}} y}{B_{c2} \xi_s} \tag{56}$$

We also assume that for narrow junctions, given the boundary conditions at the insulating surfaces and the requirement for current continuity across the S-N internal interface, $j_y^s(x)$ can be taken to be zero. Equation (56) describes the transport current density and the screening currents that flow within the junction itself. We have not included the small self field corrections to the net field, that describe the currents associated with a vortex-antivortex pair at the edges, since we assume the self-field is much smaller than the applied field.

### A. Very Narrow Junctions in High Fields

Substituting in our new expression for $\partial_x \gamma(y)$ into

$$\partial_x \left( m_x^{-1}(x) \, \partial_x f \right) + \left[ \alpha(x) - m_x^{-1} q^2 - \beta(x) f^2 - \frac{\langle j_x^s \rangle_y^2}{f^4 \, m_x^{-1}(x)} \right] f = 0 \tag{57}$$

where integrating/averaging over the $y$-direction gives $q^2 = \left( \frac{B_{\text{app}} w_s}{\sqrt{12} B_{c2} \xi_s} \right)^2$. Equation (57) represents a generalisation of Eq. (38) valid for very narrow junctions in all applied fields $B_{\text{app}}$. We can now solve for the critical current of the junction system using Eq. (57) in the two cases considered in Section IV A: when the N region is thin, when $d \ll \xi_s$; and when the N region is thick, when $d \gg \xi_s$. For all these very narrow junctions, we assume there are no vortices in the barrier.

### 1. Thin Junctions in High Fields $d \ll \xi_s$

Consider first the thin junction limit, where $d \ll \xi_s$. Assuming $\beta(x)$ and $m_x^{-1}(x)$ are constant across the system for simplicity, we rescale Eq. (57) by $\tilde{x} = x\sqrt{1-q^2}$, $\tilde{f} = f/\sqrt{1-q^2}$ and $\tilde{j}_x = \langle j_x^s \rangle_y (1-q^2)^{-3/2}$ to give

$$\partial_{\tilde{x}}^2 \tilde{f} + \left[ 1 - \frac{1-\alpha(x)}{1-q^2} - \tilde{f}^2 - \frac{\tilde{j}_x^2}{\tilde{f}^4} \right] \tilde{f} = 0 \tag{58}$$

Since $\tilde{f}$ and $\tilde{j}_x$ are continuous across the S/N interface, we find a constraint between $\partial_{\tilde{x}} \tilde{f}$ and $\tilde{f}$ at the interface in the limit where $d \ll \xi_s$, by integrating Eq. (58) across the normal region, where $|\tilde{x}| < d\sqrt{1-q^2}/2$, and assuming $\tilde{f}$ is symmetric across the junction:

$$2\tilde{f}'_{d/2} = d \frac{1-\alpha_n}{\sqrt{1-q^2}} \tilde{f}_{d/2} \tag{59}$$

where $\tilde{f}_{d/2} = \tilde{f}(x = d/2)$ and $\tilde{f}'_{d/2} = \partial_{\tilde{x}} \tilde{f}(x = d/2)$. The remainder of the derivation now follows the same approach as in Section IV A for low fields [48]; by substituting Eq. (59) into Eq. (58) and neglecting the highest order terms in the new small parameter $V_0^{-1} = \sqrt{1-q^2}/d(1-\alpha_n)$, we find the necessary condition for a solution to exist as $\tilde{j}_x < 1/2V_0$. In usual units, this corresponds to the critical current density $J_{\text{DJ}}$,

$$\lim_{d \ll \xi_s} \{ J_{\text{DJ}}(B_{\text{app}}) \} = J_0 \frac{\xi_s}{2d(1-\alpha_n)} \left( 1 - q^2 \right)^2 \tag{60}$$

where $q^2 = \left( B_{\text{app}} w_s / \sqrt{12} B_{c2} \xi_s \right)^2$ and $J_0 = B_{c2}/\kappa^2 \mu_0 \xi_s$ as before. The applied field at which the critical current density of the system is zero is given by $q^2 = 1$. This is equivalent to an applied field equal to the parallel critical field

$$B_{\text{app}}(q^2 = 1) = \frac{\sqrt{12} \xi_s}{w_s} B_{c2} \tag{61}$$

This expression has previously been found by Tinkham to be the upper critical field of a thin film superconductor of thickness $w_s$ when the applied magnetic field is parallel to the film surface, provided the film is thinner than approximately $1.8 \, \xi_s$ [51]. Equation (60) is compared to simulation data from TDGL-2D in Fig. 5, showing excellent agreement across the whole field range.

We note that the junctionless case, where $V_0 = 0$ can trivially be considered also, as the rescaling used in Eq. (58) is equivalent to rescaling the Ginzburg–Landau equations in terms of a field dependent coherence length in the superconductor $\tilde{\xi}_s = \xi_s/\sqrt{1-q^2}$. In this case, the critical current of the thin film system becomes $J_{\text{D}} (1-q^2)^{3/2}$ [51].

### 2. Thick Junctions in High Field $d \gg \xi_s$

For thick junctions, we rescale Eq. (57) into a similar form to that studied for zero-field by Fink [14] (cf Section IV A). In the superconducting regions, we rescale by $\bar{x} = x\sqrt{1-q^2}$, $\tilde{f}_s = f/\sqrt{1-q^2}$ and $\tilde{j}_x = \langle j_x^s \rangle_y (1-q^2)^{-3/2}$ to give a form equivalent to Eq. (44),

$$\partial_{\bar{x}}^2 \tilde{f}_s + \left[1 - \tilde{f}_s^2 - \frac{\tilde{j}_x^2}{\tilde{f}_s^4}\right]\tilde{f}_s = 0 \tag{62}$$

Inside the normal region, we rescale Eq. (57) by $\tilde{u} = x\sqrt{\frac{m_n}{m_s}\left(-\alpha_n + \frac{m_s}{m_n}q^2\right)}$, $\tilde{f}_n = -f\sqrt{\beta_n/\left(-\alpha_n + \frac{m_s}{m_n}q^2\right)}$ and $\tilde{j}_u = \langle j_x^s \rangle_y \beta_n \sqrt{m_n/m_s}(-\alpha_n + \frac{m_s}{m_n}q^2)^{-3/2}$ to give a form similar to Eq. (45),

$$-\partial_{\tilde{u}}^2 \tilde{f}_n + \left[1 - \tilde{f}_n^2 + \frac{\tilde{j}_u^2}{\tilde{f}_n^4}\right]\tilde{f}_n = 0 \tag{63}$$

The critical current in field can now be obtained following the procedure used by [14] for zero field, but with the new, field-dependent rescaled variables. In usual units, the critical current of this narrow junction system in applied fields is given by:

$$\lim_{d \gg \xi_s > w_s}\{J_{\text{DJ}}(B_{\text{app}})\} = 4J_0(1-q^2)^{\frac{3}{2}}\frac{1-\sqrt{1-\tilde{s}\tilde{f}_{d/2}^2}}{\tilde{s}\tilde{v}}\exp\left(-\frac{d}{\tilde{\xi}_n}\right), \tag{64}$$

where

$$\tilde{f}_{d/2}^2 = \frac{\tilde{v}^2 + 1 - \sqrt{\tilde{v}^2(2-\tilde{s})+1}}{\tilde{v}^2 + \tilde{s}}, \qquad \tilde{v} = \frac{m_n \tilde{\xi}_n}{m_s \xi_s}\sqrt{1-q^2},$$

$$q^2 = \frac{B_{\text{app}}^2 w_s^2}{12}, \qquad \tilde{s} = \frac{\beta_n(1-q^2)}{(\alpha_n - \frac{m_s}{m_n}q^2)}, \qquad \tilde{\xi}_n = \sqrt{\frac{m_s}{m_n}\frac{1}{\left(-\alpha_n + \frac{m_s}{m_n}q^2\right)}}\xi_s \tag{65}$$

and $J_0 = B_{c2}/\kappa^2\mu_0\xi_s$ as in Section II. Once again, here we take $\beta_n = 1$ and so when the effective mass of the N region is the same as that of the superconductors, $\tilde{v}^2 \to -\tilde{s}$, and $\tilde{f}_{d/2}^2 \to (1-q^2)/2(1-\alpha_n)$. Equation (64) is compared to the critical current densities obtained from TDGL-2D in Fig. 6. Excellent agreement between Eq. (64) and TDGL-2D is observed across the entire field range, and across the parameter space for $d > \xi_s$, $\alpha_n < -1.0$, and $0.1\,m_s < m_n < 6.0\,m_s$.

In the limit where $\tilde{f}_{d/2}^2 \to 0$, and when $m_n = m_s$ Eq. (64) reduces to the simpler form

$$\lim_{d \gg \xi_s > w_s}\{J_{\text{DJ}}(B_{\text{app}})\} = J_0 \frac{(1-q^2)^2}{\sqrt{1-\alpha_n}}\exp\left(-\frac{d\sqrt{1-\alpha_n}}{\xi_s}\right) \tag{66}$$

which provides the general field-dependent form for Eq. (48) famously found by De Gennes for SNS junctions in zero field [12]. In general, weakly coupled junctions with $\tilde{f}_{d/2}^2 \to 0$ for any thickness of junction with $m_n = m_s$ can be described by the single expression

$$\lim_{\xi_s > w_s}\{J_{\text{DJ}}(B_{\text{app}})\} = J_0 \frac{(1-q^2)^2}{2\sqrt{1-\alpha_n}\sinh\left(d\sqrt{1-\alpha_n}/\xi_s\right)} \tag{67}$$

where Eq. (60) is recovered in the limit $d\sqrt{1-\alpha_n}/\xi_s \to 0$ and Eq. (66) is recovered in the limit $d\sqrt{1-\alpha_n}/\xi_s \gg 1$.

### 3. Comments and comparisons

The new solutions derived in this work for very narrow junctions, Eqs. (60) and (64), are formally restricted to systems with weakly coupled junctions with width of order of the coherence length, bounded by insulating surfaces. In this regime, an increase in magnetic field induces large screening currents in the superconductor close to the junction, which are restricted to flow parallel to the film surfaces due to the insulating boundary conditions and weaken superconductivity in the film. In effect, the applied magnetic field acts to increase the energy of the superconducting state, making it less stable, and increasing the local coherence length in the superconductor (and decreasing it in the normal metal). This reduces the magnitude of the order parameter far from the junction and increases the length scale over which the order parameter recovers from the boundary with the normal metal. At the parallel critical field, this length diverges, and superconductivity is destroyed throughout the system.

The full-field approximation for $J_c$ given in Eq. (64) has the same leading order monotonically decreasing behaviour in low field as predicted by the authors of [22, 23, 52] using a model of an SNS Josephson junction from the linearised Usadel equations, including the applied magnetic field as an effective spin-flip scattering rate. Indeed, our result, Eq. (64) can be viewed as an extension to this result that describes fields approaching the parallel critical field of the superconductor.

Experimental measurements of SNS junctions consisting of superconducting nanowires in this monotonically decaying regime that have been carried out in [53, 54] show good agreement with Eq. (64), as shown in Fig. 7 with reasonable estimates for the coherence length in the superconducting nanowires.

### B. Narrow Junctions

We now extend our new solutions for $J_c(B_{app})$ in very narrow junctions to describe the qualitative behaviour of larger 2D systems with narrow junctions, with widths up to the length scale of the superconductor penetration depth $\lambda_s$, in arbitrary applied magnetic fields. In low fields, Eq. (54) accounts for the fraction of the total width of the junction over which current density flows, as a result of screening currents in the superconductor set up by the distribution of vortices inside the junction. We therefore expect any approximation to $J_c(B_{app})$ to reduce to this expression when the applied field is far below the critical magnetic field of the junction. In high fields, both the order parameter and the local current density vary on a length scale of order the vortex-vortex spacing $a_0$, instead of the junction width $w_s$. We therefore replace the zero field $J_{DJ}$ term in Eq. (54) with the field dependent $J_{DJ}$ expressions from Eqs. (43) and (64) but with the width $w_s$ replaced by a term comparable to the vortex spacing in the superconductor $\sim a_0$. This yields our approximation for $J_c$ for narrow junctions over the full field range as

$$J_c(B_{app}) = c_0 \left(\frac{\phi_0}{B_{app} w_s^2}\right)^{c_1} J_{DJ}\left(B_{app}, w_s \to a_0\right), \quad (68)$$

where we set $q^2 = B_{app}/B_{c2}^*$ and $J_{DJ}$ is taken from Eq. (60) and Eq. (64) in the thin limit and in the thick limit respectively. We have replaced $B_{c2}$ by $B_{c2}^*$ to include junctions such as that considered above, where there is a insulating surface barrier along the edge of both the superconductor and the junction. This ensures that in the case of a simple thin film between two insulators, the result $J_c \approx J_D \left(1 - B_{app}/B_{c2}^*\right)^{3/2}$ is obtained, as found previously by Abrikosov [55] close to the upper critical field of the system. Explicitly, in the weak coupling limit, Eq. (68) for thin junctions takes the form,

$$J_{\rm c}(B_{\rm app}) = J_0 \frac{c_0 \xi_{\rm s}}{2d(1-\alpha_{\rm n})} \left(\frac{\phi_0}{B_{\rm app} w_{\rm s}^2}\right)^{c_1} \left(1 - \frac{B_{\rm app}}{B_{c2}^*}\right)^2, \qquad (69)$$

whereas for thick junctions,

$$J_{\rm c}(B_{\rm app}) = J_0 \frac{c_0}{\sqrt{1-\alpha_{\rm n}}} \exp\left(-\frac{d\sqrt{1-\alpha_{\rm n}}}{\xi_{\rm s}}\right) \left(\frac{\phi_0}{B_{\rm app} w_{\rm s}^2}\right)^{c_1} \left(1 - \frac{B_{\rm app}}{B_{c2}^*}\right)^2. \qquad (70)$$

2D simulations for two narrow junctions in high field are plotted in Fig. 8 and compared to Eq. (69) with $c_0 = c_1 = 0.58$ and $B_{c2}^*$ set to $1.8 B_{c2}$. Excellent agreement is seen between the analytic functional form and the simulated data, with only $B_{c2}^*$ taken as a free parameter. We note that the power law dependence with $c_1 \approx 0.6$ has also been widely observed in many high temperature superconductors at high temperatures and magnetic fields that, as expected, are well below $B_{c2}^*$ [34, 56, 57]

## VI. 3D POLYCRYSTAL FLUX FLOW AND CRITICAL CURRENT SIMULATIONS

The morphology of grain boundaries in real 3D systems is significantly more complex than that considered in the 2D Josephson junction simulations of Section V. Here we investigate the critical current density that can be carried by a 3D polycrystalline system containing Josephson junction-like grain boundaries using the TDGL-HI$\kappa$ algorithm [35].

### A. Polycrystalline Simulations

To create our model polycrystalline material for critical current and flux pinning simulations, we first generate a 3D tessellation of equiaxed grains, periodic in all three dimensions, with grain sizes corresponding to a typical lognormal grain size distribution for a grain growth system, using the *Neper* software package v3.5.0 [58, 59]. In this package, the size of a grain is represented by the effective diameter $D$ of a spherical cell with an equivalent volume to the grain. The grain size $D$ and grain sphericity $s$ distributions are controllable; for a typical grain-growth polycrystal, whose grain boundaries have migrated during formation from capillarity effects, the grain size distribution $D/\langle D \rangle$ has been observed to follow a lognormal distribution with average 1 and standard deviation 0.35, and $1-s$ follows a lognormal distribution of average 0.145 and standard deviation 0.030 [59]. The resultant grain size distributions are broader and generate more spherical grains than those generated by a 3D Voronoi tessellation on the same domain. For use as a simulation output, this tessellation is postprocessed, with every mesh point in the superconducting volume within a distance $d/2$ of a face of a crystal grain assigned grain boundary properties with $\alpha = \alpha_{\rm GB}$. In this manner, a rasterised approximation to an equiaxed polycrystal is constructed, with grain boundaries given degraded superconducting properties with $\alpha_{\rm GB} < 1$.

The base parameters of our model polycrystalline system are given in Table I. We consider Nb$_3$Sn at $T = 4.2$ K with a critical temperature of $T_{\rm cs} = 17.8$ K and a coherence length $\xi_{\rm s}(4.2\text{ K}) \approx 3.12$ nm. The superconducting volume of our base system corresponds to physical dimensions of 468 nm × 468 nm × 468 nm with a mean grain size $D = 70$ nm. An example distribution of grain boundaries for this set of parameters, along with distributions of $|\psi|$ over the simulation domain and close to a representative grain are presented in Fig. 9.

The flux pinning force per unit volume $F_{\rm p} = J_{\rm c} B_{\rm app}$ as a function of reduced field, for polycrystalline material with different grain boundary parameters $\alpha_{\rm GB}$, obtained from TDGL-HI$\kappa$, are shown in the upper panel of Fig. 10. The optimum flux pinning forces occur when the grain boundary thickness $d_{GB}$ is close to the effective (normal metal) coherence length in the grain boundary $\xi_{GB} = \sqrt{-\alpha_{GB}}\,\xi_s$ (defined when $\alpha_{GB} < 0$). For more degraded boundaries, $J_c$ decays approximately exponentially at a rate proportional to $d_{GB}/\xi_{GB}$ for $d_{GB}/\xi_{GB} > 1$ and for $\alpha_{GB} < -4.0$ the maximum in the flux pinning force $F_p \propto J_c B_{app}$ is found at higher reduced field values. For more weakly degraded grain boundaries ($\alpha_{GB} > -4.0$), we find a Kramer dependence [29, 60] such that the maximum flux pinning force per unit volume is close to 0.2 $B_{c2}$ and consistent with the field dependence of other computational results obtained using a different polycrystalline grain morphology [2]. Both the magnitude of $J_c$ with a grain size of 70 nm at $10^{-3}J_D$, and the Kramer field dependence, are similar to those observed experimentally in optimised polycrystalline Nb$_3$Sn [1] suggesting the simulations capture the important physical processes in these systems. In the time dependent simulations when $J > J_c$ (i.e. showing continuous vortex movement), we see significant differences in the curvature of moving vortices, above and below the optimum. In strongly degraded boundaries when $\alpha_{GB} < -4.0$, vortices are significantly curved and follow grain boundaries, being preferentially held at points where two or more grain boundaries meet, whereas for $\alpha_{GB} > -2.0$, vortices remain mostly straight, aligned along the applied field in the $z$ axis. Experimental and simulation flux pinning curves for different mean grain sizes are presented and compared in the lower panel of Fig. 11. For consistency, we have confirmed that in homogeneous systems with no flux pinning structures present, no significant critical current densities are found in these simulations. The maximum flux pinning force per unit volume as a function of grain size is similar to the experimental values. These results enable us to tune grain boundary properties and morphologies that provide directions for better polycrystalline materials. For very small grain sizes with $D < 100$ nm, our simulations show $F_p^{max}$ values that are larger than observed in experiment, and this can be attributed to a shorter coherence length in the superconducting grains $\xi_s$ for these fine grained systems or degraded grain boundaries in such small grain material. To our knowledge, these simulations are the first for 3D polycrystalline systems that display this increase of $F_p^{max}$ with decreasing grain size $D$ in qualitative agreement with experiment.

### B. Flux Pinning in Polycrystalline Materials

The Kramer-like field dependence implied by Eq. (69) has been widely observed in low temperature polycrystalline superconductors such as Nb$_3$Sn [31] up to $B_{c2}$, and the $w^{-1.2}$ factor in Eq. (68) is reminiscent of the inverse grain size dependence observed for $J_c$ experimentally [61] and in our simulations (Fig. 11). Motivated by this, we propose an expression for the flux pinning force per unit volume for a polycrystalline system with weakly coupled grains (with highly degraded grain boundaries) based on Eq. (68) in order to relate the results here to standard flux pinning formulism, where

$$F_p(B_{app}) \approx J_0 B_{c2} A \left(\frac{\phi_0}{B_{c2}^* D^2}\right)^r (b^*)^p (1 - b^*)^q f(\alpha_{GB}), \tag{71}$$

and we have replaced $w_s$ by the grain size $D$; defined the pinning parameters $p \approx 1 - c_1$ and $q \approx 2$; introduced the new empirical parameters $A$ and $r$; and made the weak coupling approximations that $f(\alpha_{GB}) = \xi_s/2d\,(1 - \alpha_{GB})$ in the thin limit and $f(\alpha_{GB}) = \exp\left(-d\sqrt{1-\alpha_{GB}}/\xi_s\right)/\sqrt{1-\alpha_{GB}}$ in the thick junction limit for the grain boundary (GB). $F_p^{max}$ is found as usual at the field $b^* = p/(p+q)$. The empirical parameters $A$ and $r$ account for the fraction of the total vortex length that is held within grain boundaries.

Comparisons of Eq. (71) in the thick junction limit to our TDGL results are presented in the upper panels of Figs. 10 and 11. $A$, $p$ and $q$ were taken to be free parameters for each flux pinning curve, and $r = 1.1$ was obtained as a global fit parameter from the combined set of simulations. The maximum in the flux pinning force per unit volume, $F_\text{p}^\text{max}$, has been compared to a constrained form of Eq. (71) in the lower panels of Figs. 10 and 11, in which the pinning parameters are restricted to their Kramer-like values $p = 0.5$, $q = 2$. The decrease in critical current density as the grain boundary properties degrade (as $\sqrt{1 - \alpha_\text{GB}}$ increases) in the weak coupling limit of grains appears to be well represented by Eq. (71) and $f(\alpha_\text{GB})$ taken from Eq. (66). In this case, the parameters $A$ and $r$ are closely related to their 2D equivalents in Eq. (70), with $r \approx c_1 \approx 0.6$ and in the limit of strongly degraded grain boundaries, $A \approx c_0/3$, as shown by Fig. 10. The observation that the prefactor $c_0$ in the 2D junction simulations is approximately three times larger than the prefactor $A$ in the 3D simulations here may partly be due to the stronger surface barrier existing in the junction system at the junction-insulator interface. The surface barrier at the grain-grain boundary interface in the 3D simulations is generally weaker as a result of the proximity effect limiting supercurrents at the interface, similar to the effect observed at metallic interfaces. For the polycrystal system in Table I, which lies close to the peak $F_\text{p,max}$ in Fig. 11, $J_\text{c} \sim b^{-0.4}(1-b)^{2.7}$ ($p = 0.6$, $q = 2.7$), close to the Kramer-like field dependence of the critical current density $J_\text{c} \sim b^{-0.5}(1-b)^2$ ($p = 0.5$, $q = 2$). Deviations of $p$ and $q$ from predictions can occur due to multiple pinning mechanisms contributing to $J_\text{c}$ concurrently; indeed, videos of the simulated vortex state in motion (not shown here) show complex vortex depinning from grain boundaries, line intersections, and triple points across the range of $\alpha_\text{GB}$ in Fig. 10.

## VII. DISCUSSION AND CONCLUSIONS

It is important to note that all the polycrystalline simulations carried out in this work are in the high-$\kappa$ limit, when the local magnetic field is equal to the applied magnetic field in the system at every point. Nevertheless, we expect the results to be qualitatively accurate for real systems of materials such as $Nb_3Sn$, since the penetration depth in such materials $\lambda_\text{s} \approx 100\,\text{nm}$ is still of the order of the grain size [1], and so in high fields, magnetisation of grains will still be small relative to the applied magnetic field. The same is not necessarily true in very weak applied fields though, and thus care should be taken interpreting results in weak applied fields as a result. Nevertheless, large-scale TDGL simulations provide a useful possible tool for the study of how changes in grain structure and the grain boundary network in polycrystalline materials affect the critical current density that the material can carry, and for visualising the manner in which vortices flow in such materials close to the critical current. Fabrication and measurement of real samples displaying systematic variations in grain size can be difficult and time-consuming to achieve experimentally, but investigation of grain size effects in simulated systems can be performed in much shorter time frames. In this work, we have obtained new expressions for the critical current density of narrow, tunnel-like SNS Josephson junctions on the scale of the superconducting coherence length across the entire magnetic field range, up to the effective upper critical field of the superconducting system. To the best of our knowledge, these expressions for $J_\text{c}$ are the first for any Josephson junction system to be valid up to the effective upper critical magnetic field. We have confirmed these expressions against simulations based on time-dependent Ginzburg–Landau theory, and validated existing expressions from the literature for the critical current of Josephson junctions in low applied magnetic fields. We have also found these expressions to be consistent with experimental data for W-Au nanowire junctions [53] and Al-Au nanowire junctions [54] in which monotonically decreasing critical currents with field have been observed. By applying these new expressions for the critical current density of narrow junctions to the edge regions of Josephson junctions with dimensions much larger than the superconducting coherence length but smaller than the superconductor penetration depth, we obtain expressions for the critical current density as a function of field from a junction-based model that qualitatively agrees with experimental data for polycrystalline superconductors such as $Nb_3Sn$ and existing models based on flux shear through grain boundaries [61]. We have also performed 3D simulations of equiaxed polycrystalline systems in the high $\kappa$ limit, which we believe to be the first for a complex polycrystalline system to show an increase in the critical current density of the system with decreasing grain size in qualitative agreement with experiment [62]. Such simulations display maximum critical currents when the grain boundary thickness is similar to the effective coherence length in the grain boundary regions, at which point the field dependence of the critical current density $J_c \approx b^{1/2}(1-b)^2$. As a consequence of these simulations, we conclude that at the critical current of polycrystalline superconducting materials optimised for high field applications, the dissipative state can usefully be described as vortex flow along grain boundaries, with flux pinning from the lines at which grain boundaries intersect one another.

**VIII. ACKNOWLEDGEMENTS**

This work is funded by EPSRC grant EP/L01663X/1 that supports the EPSRC Centre for Doctoral Training in the Science and Technology of Fusion Energy. This work has been carried out within the framework of the EUROfusion Consortium and has received funding from the Euratom research and training programme 2014-2018 under grant agreement No. 633053. Data are available at xxxx and associated materials are on the Durham Research Online website: http://dro.dur.ac.uk/.

This work made use of the facilities of the Hamilton HPC Service of Durham University. The authors would like to thank M. Raine, A. Smith, J. Greenwood, S. Chislett-McDonald, C. Gurnham, B. Din and P. Branch for useful discussions. One of the authors (DPH) would like to acknowledge the many tutorials provided for him over the years at conferences, the discussions and the fabulous generosity of spirit from John R. Clem (1938-2013) and Archie Campbell (1940-2019). The code is available on request from DPH.

**FIGURES**

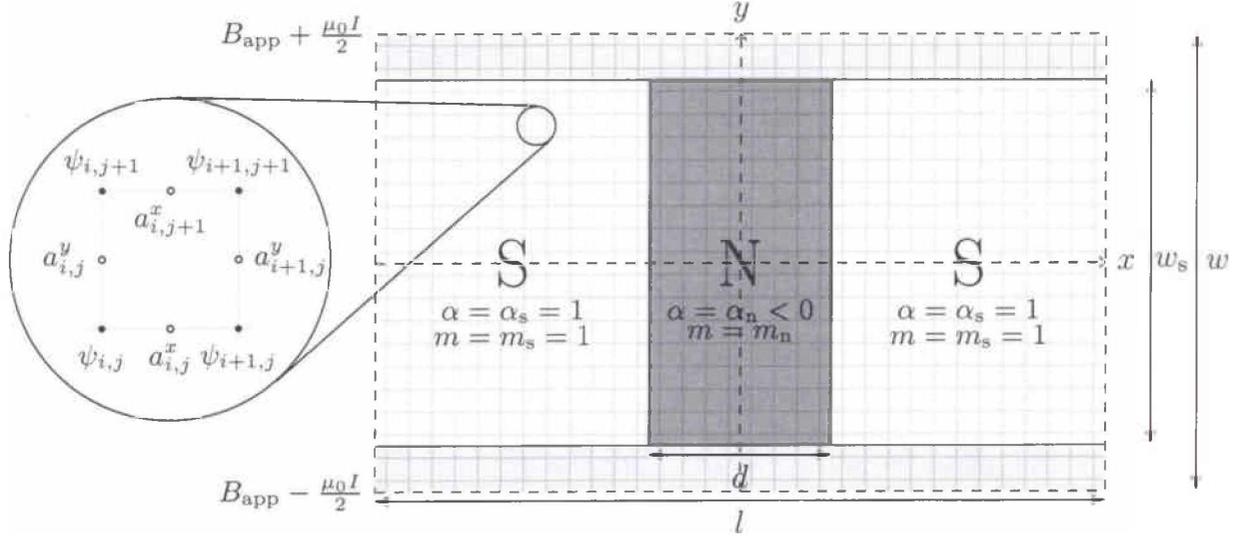

FIG. 1: Schematic of the 2D computational domain of width $w$ and periodic length $l$ used to model the junction system. The domain is subdivided into three sections; the main superconducting region, S, in which the normalised Ginzburg–Landau temperature parameter $\alpha = 1$ and normalised effective mass $m = 1$, a normal region N described by the normalised Ginzburg–Landau temperature parameter and effective mass $\alpha_n$ and $m_n$ respectively, and a coating region, marked in light grey, in which $\alpha = -10.0$ and $m = 10^8$ when modelling junctions with insulating coatings. The applied field $B_{\mathrm{app}}$ and current $I$ are controlled by fixing the local magnetic field at the edges of the computational domain in the $y$ direction. The junction thickness in the direction of current flow is denoted $d$ and the junction width is denoted $w_s$ Exploded view: schematic of the location at which the discretised order parameter $\psi_{i,j}$ and modified link variables $a^x_{i,j}$ and $a^y_{i,j}$ relative to the underlying computational grid. Unless otherwise stated, the grid step size is typically taken to be $h_x = h_y = 0.5\xi_s$ in these simulations.

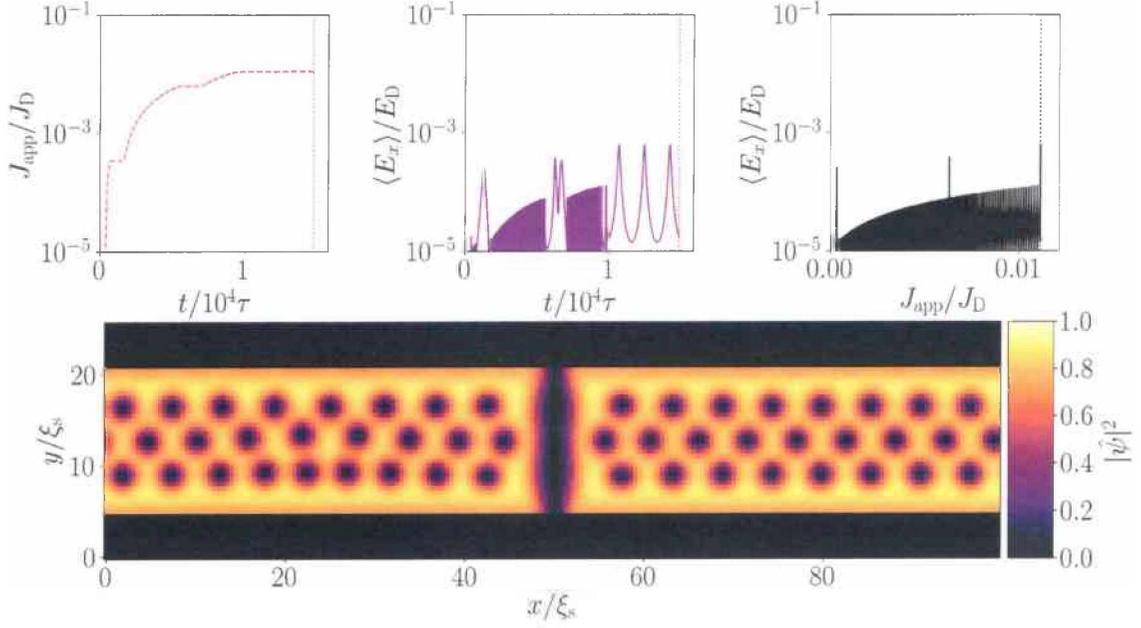

FIG. 2: Typical simulation data used to extract $J_c$ at the applied field $B_{\text{app}} = 0.3 B_{c2}$. Bottom: distribution of the normalised Cooper pair density $|\tilde{\psi}|^2$ at the critical current $J_c$, for a simulated junction with periodic length $l = 100\xi_s$, thickness $d = 0.5\xi_s$, junction width $w_s = 16.0\xi_s$ and Ginzburg–Landau temperature parameter in the normal region $\alpha_n = -20$. Top left: The applied current density $J_{\text{app}}$ normalised by the depairing current density $J_D$ versus time $t$ normalised in units of the characteristic timescale $\tau$. Top centre: The average electric field in the $x$ direction $\langle E_x \rangle$ normalised by the characteristic electric field $E_D$ as a function of time $t$. Top right: The normalised average electric field in the $x$ direction as a function of the applied current density. The applied current density when $E < E_c = 10^{-5} E_D$, and $J_c$ is determined as the lowest current at which $E > E_c$ for a duration exceeding $t_{\text{hold}} = 5 \times 10^3 \tau$.

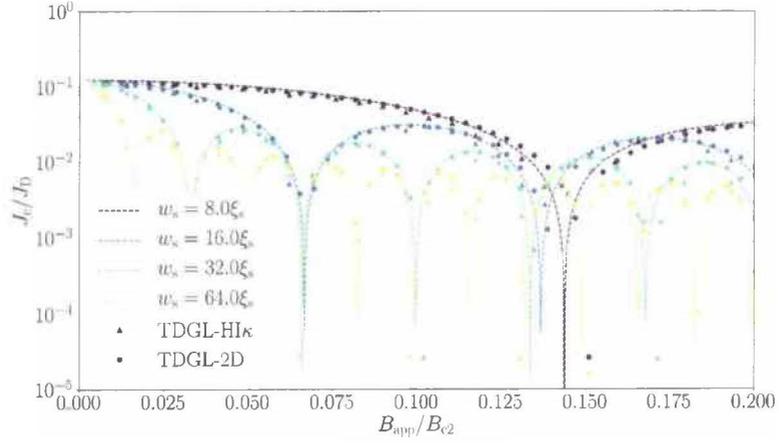

(a)

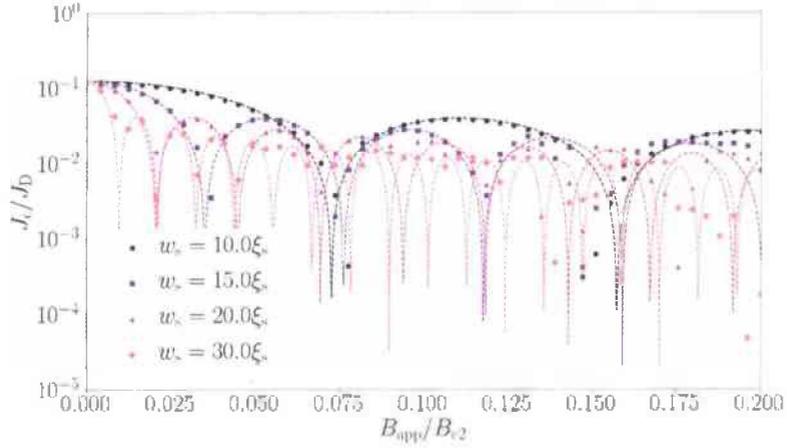

(b)

FIG. 3: Simulations of $J_c(B)$ of narrow, very thin, weakly coupled junctions with different widths $w_s$. The system size in the $x$-direction $l = 6.0\ \xi_s$ (Upper) and $100.0\ \xi_s$ (Lower). The junction thickness $d$ was taken to be $d_{\min} = 0.5\ \xi_s$, $\alpha_n = -20.0$ and $\kappa = 40.0$. Top: $J_c(B)$ as calculated using the TDGL-2D code (circles) and TDGL-HI$\kappa$ code (triangles), with the hold time and time step for the TDGL-2D simulations set to $t_{\text{hold}} = 5 \times 10^3 \tau$ and $\delta t = 0.5\tau$, and for the TDGL-HI$\kappa$ simulations set to $t_{\text{hold}} = 10\tau$ and $\delta t = 0.1\tau$ respectively. Bottom: $J_c(B)$ as calculated using the TDGL-2D code with hold time $t_{\text{hold}} = 10^3 \tau$ and time step $0.1\tau$. Dashed lines in both panels are given by Eqs. (50) and (51) with $d_{\text{eff}} = 2\xi_s$.

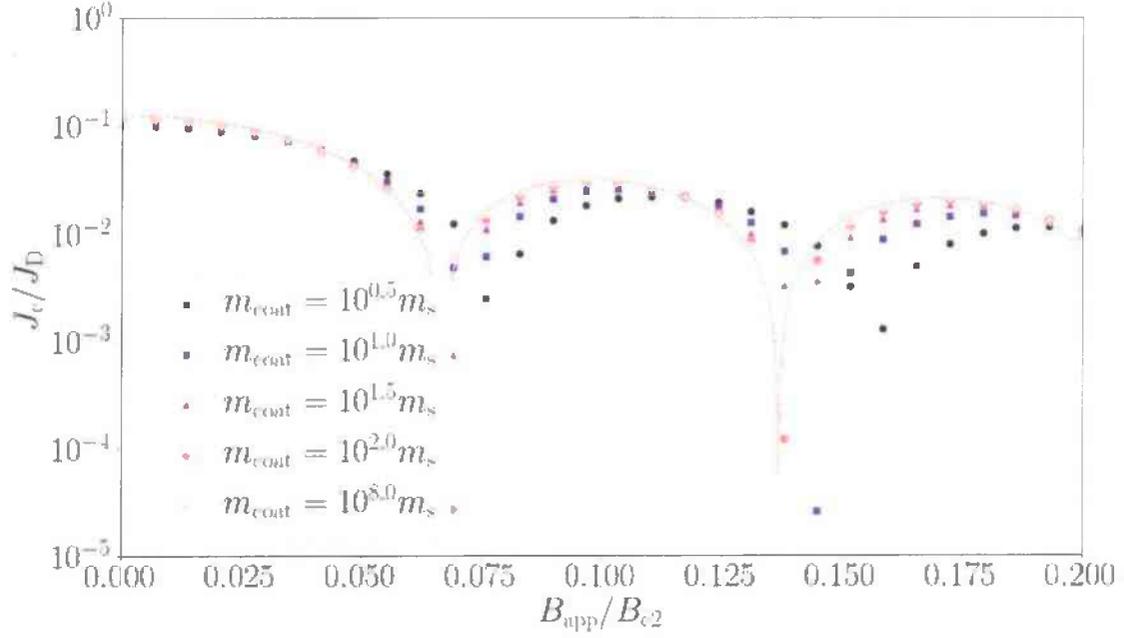

FIG. 4: Simulations of the critical current of a very thin junction in the weak coupling limit with the Ginzburg–Landau temperature parameter in the normal region $\alpha_n = -20.0$, a junction thickness $d = 0.5\xi_s$ smaller than the superconducting coherence length $\xi_s$, and a width $w_s = 16\xi_s$ much smaller than the Josephson penetration depth $\lambda_J$ for varying coating effective mass (proportional to the coating resistivity) with a coating thickness of $5\xi_s$. The periodic system size in the $x$-direction $l = 6.0\xi_s$, and the Ginzburg–Landau parameter and friction coefficient in the superconductor are $\kappa = 40.0$ and $\eta = 5.79$ respectively throughout. For this system, coating masses below $\sim 30 m_s$ show distortion of the Fraunhöfer pattern, with reduced zero field $J_c$ and increased spacing between minima in the $J_c$ characteristic relative to the insulating coating limit ($m_{\text{coat}} \to \infty$) Remaining computational parameters are as described in the text. Dashed lines in both panels are given by Eqs. (50) and (51) with $d_{\text{eff}} = 2\xi_s$.

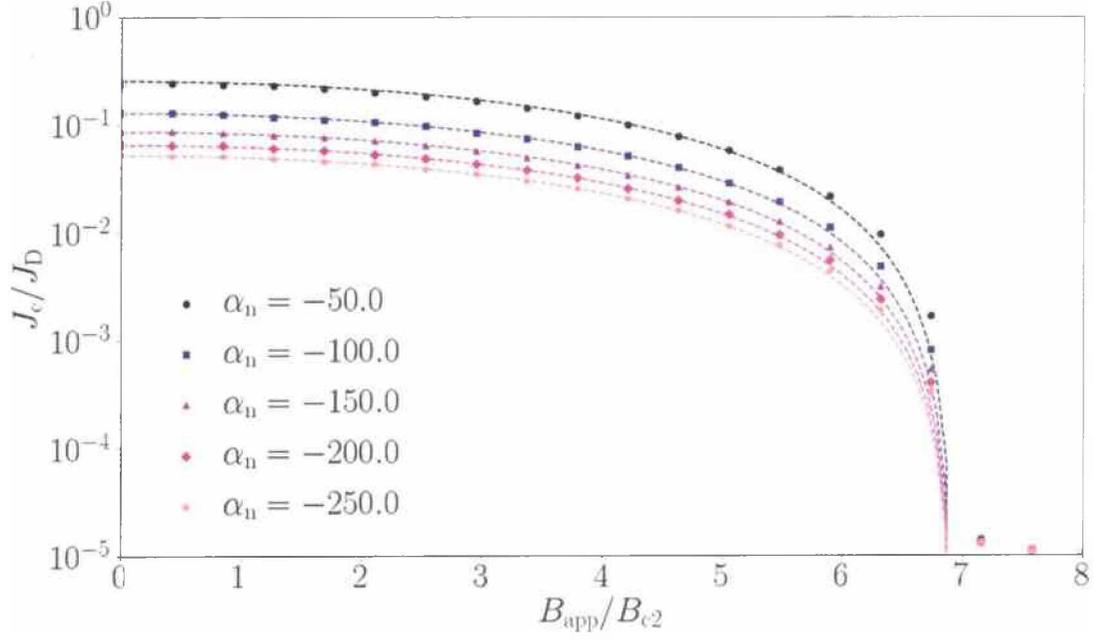

FIG. 5: Simulations of $J_c(B)$ of very narrow, thin, weakly coupled junctions as a function of $\alpha_n$ where $-250 \leq \alpha_n \leq -50$. The width $w_s = 0.5\xi_s$ and the junction thickness $d = d_{\min} = 0.1\xi_s$. The periodic system length in the $x$ direction $l = 12.0\xi_s$ and $\kappa = 5$. The effective mass in the normal region was taken to be $m_n = m_s$. The grid spacing was chosen to be $h_x = h_y = 0.1\xi_s$, the time step $\delta t = 0.5\tau$, and the hold time $t_{\text{hold}} = 5 \times 10^3 \tau$. Dashed lines are given by Eq. (43).

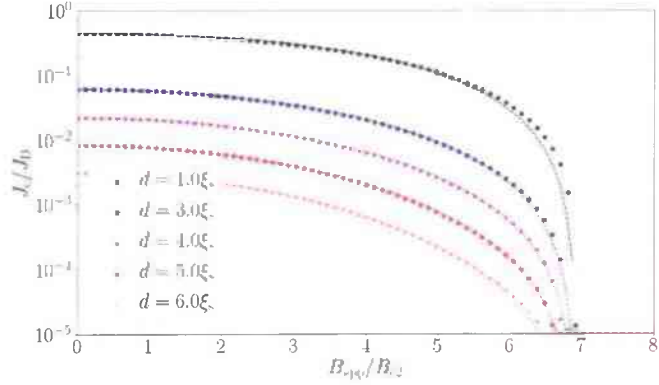

(a)

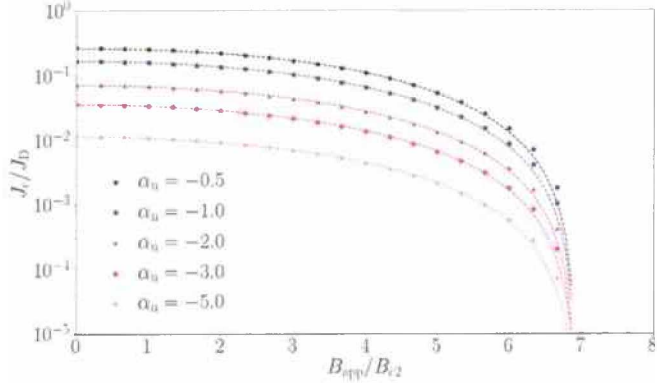

(b)

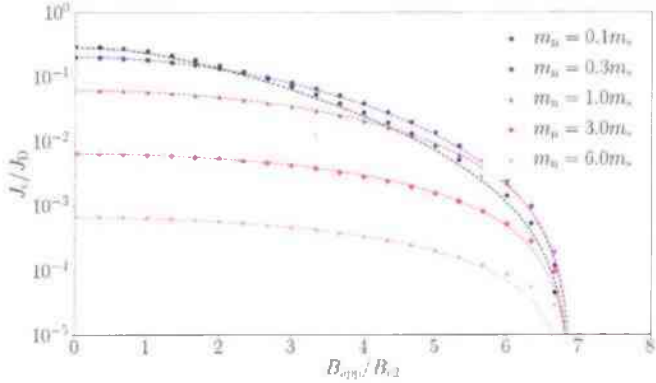

(c)

FIG. 6: Simulations of $J_c(B)$ for very narrow, thick, weakly coupled junctions. The width $w_s = 0.5\xi_s$, the periodic system length in the $x$ direction $l = 12.0\xi_s$ and $\kappa = 5$. The grid spacing was $h_x = h_y = 0.1\xi_s$, the time step $\delta t = 0.5\tau$, and the hold time $t_{\text{hold}} = 5 \times 10^3 \tau$. (Upper) The effective mass in the normal region was taken to be $m_n = m_s$, $\alpha_n = -1.0$, and the junction thickness $d$ was varied. (Middle) $m_n = m_s$, $\alpha_n$ was varied and $d = 2.0\xi_s$. (Lower) $m_n$ was varied, $\alpha_n = -1.0$ and $d = 2.0\xi_s$. Dashed lines in all panels are given by Eq. (64).

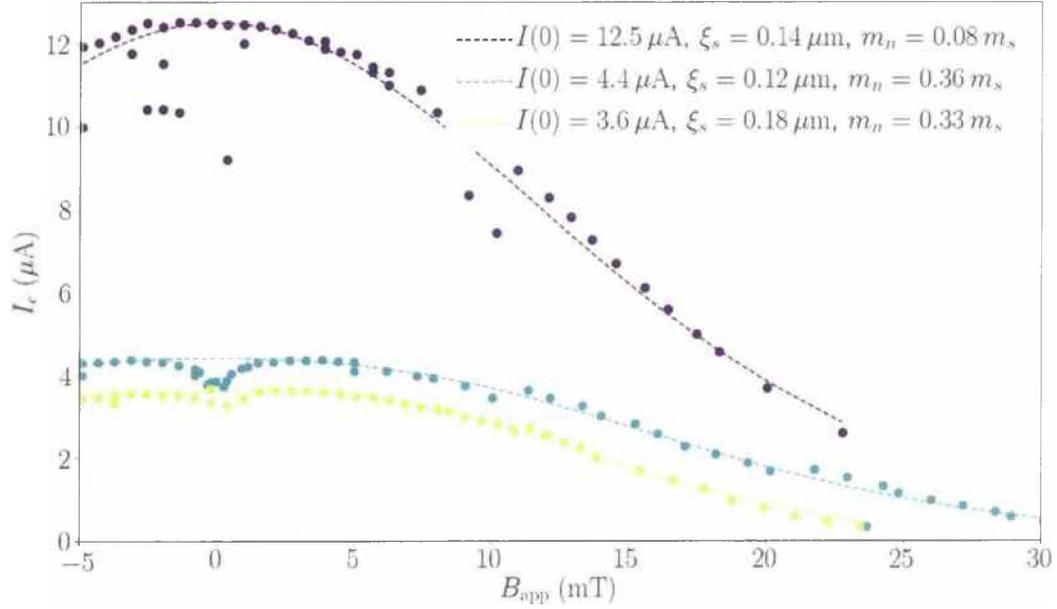

FIG. 7: Comparison of Eq. (64) to experimental data on Al-Au-Al nanowire junctions measured by [54]. The junction thickness $d$ varied between 900 and 1300 nm, and all junctions were $w_s = 125$ nm wide. The coherence length $\xi_n$ in the Au region was taken to be 10 µm as suggested by weak localization experiments below 50 mK. The critical current at zero field $I(0)$ was fixed at the maximum measured current, and the coherence length of the Al superconductor $\xi_s$ along with the ratio of the effective mass of a Cooper pair in Au and in Al $m_n/m_s$ were left as free parameters for the fit.

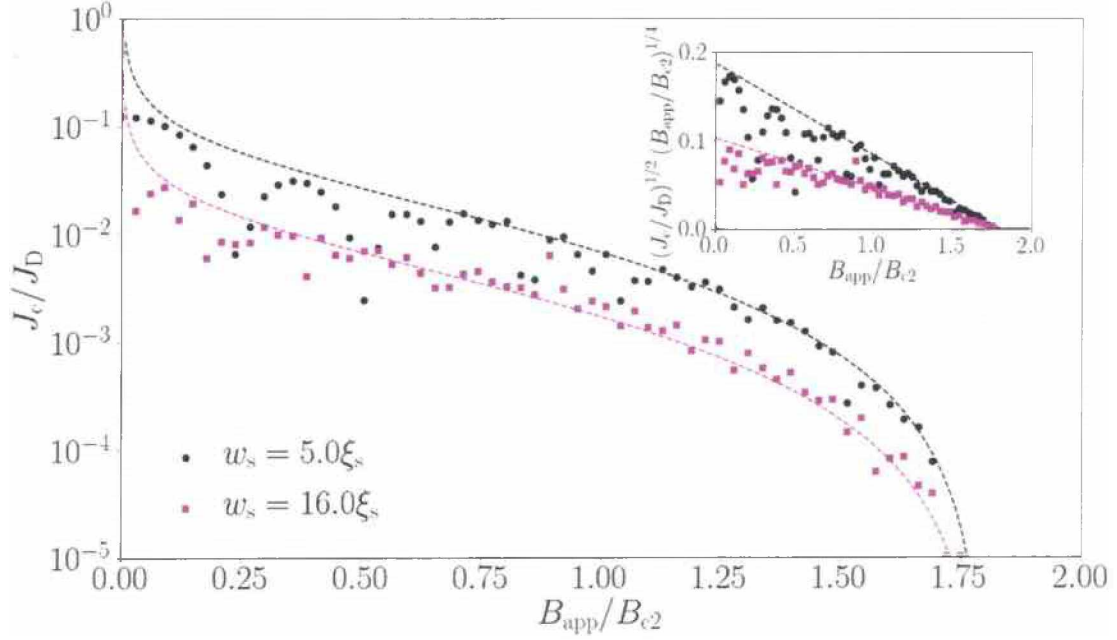

FIG. 8: Simulations of the critical current of a narrow, thin junction in the weak coupling limit (markers) as described in Section IV B with the Ginzburg–Landau temperature parameter in the normal region $\alpha_n = -40.0$, a junction thickness $d = 0.25\xi_s$ smaller than the superconducting coherence length $\xi_s$, and a width $w_s$ much smaller than the Josephson penetration depth $\lambda_J$ but much larger than $\xi_s$. The periodic system size in the $x$-direction $l = 100.0\xi_s$, and the Ginzburg–Landau parameter and friction coefficient in the superconductor are $\kappa = 40.0$ and $\eta = 5.79$ respectively throughout. The grid spacing was chosen to be $h_x = h_y = 0.25\xi_s$ and the time step $\delta t = 0.5\tau$. Dashed lines represent Eq. (69) for the example parameters $B_{c2}^* = 1.8B_{c2}$ and $c_0 = c_1 = 0.58$. Remaining computational parameters are as described in the text. Inset: Kramer plot of data shown in main plot.

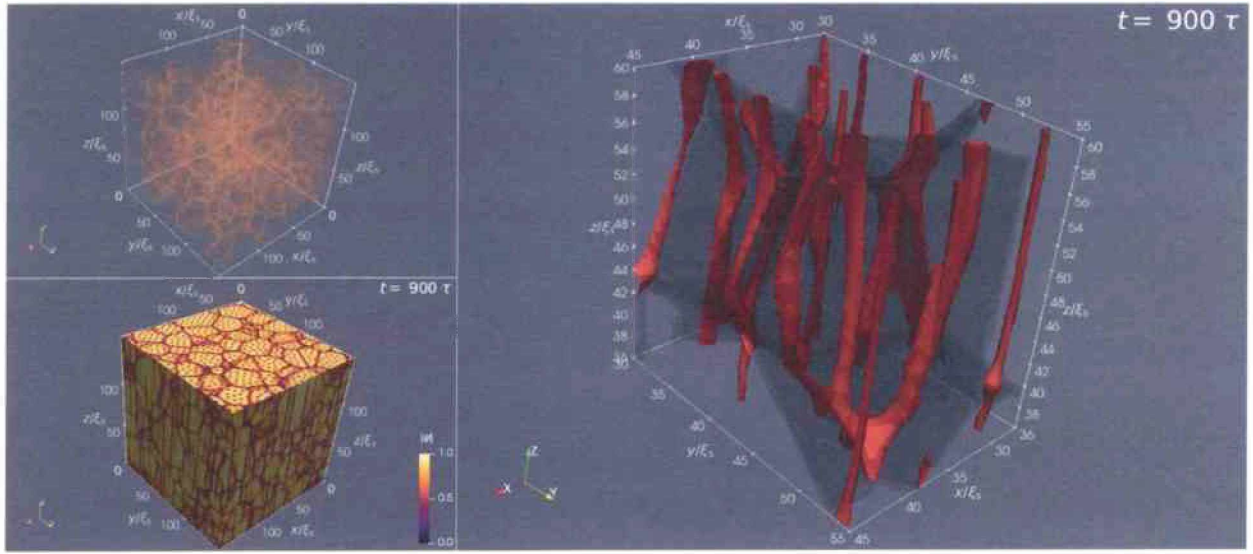

FIG. 9: A snapshot of the time dependent simulation at $J_{\text{app}} = 10^{-2} J_{\text{D}}$ and $B_{\text{app}} = 0.2 B_{\text{c2}}$ for the base system described in Table I. Top left: grain boundary network of the periodic physical system. Bottom right: distribution of the magnitude of the order parameter $|\psi|$ across the surfaces of the computational domain. Right: distribution of vortices around an example grain in the system. The surface of the region enclosing points where $|\psi| < 0.25$ is displayed in red, and the grain boundary regions are shown in black.

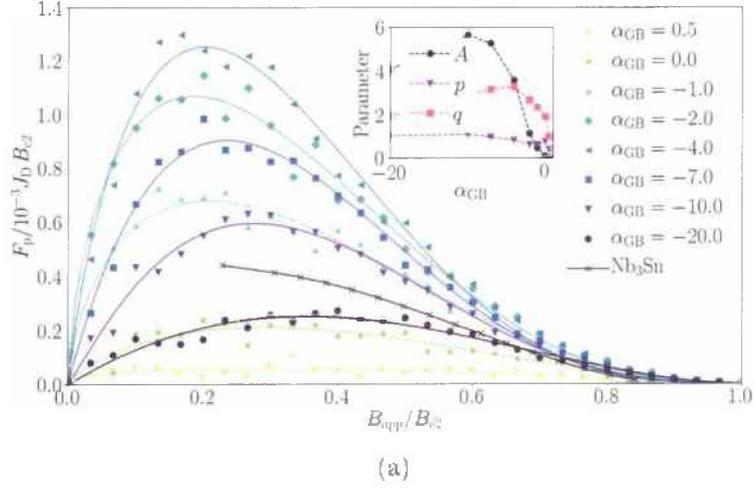

(a)

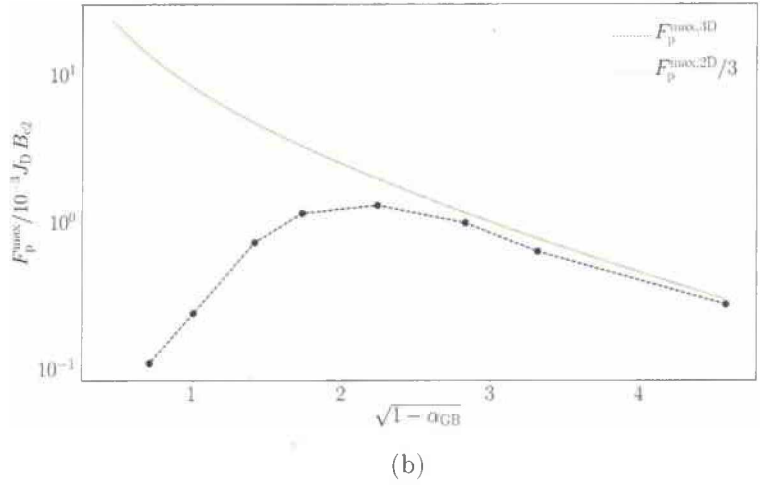

(b)

FIG. 10: (Upper): Normalised flux pinning force $F_p/10^{-3} J_D B_{c2}$ for the polycrystalline 3D system described in Table I with varying $\alpha_{GB}$ at various applied magnetic fields. The maximum in the flux pinning force is found close to $B_{app} = 0.2 B_{c2}$ for $\alpha_{GB} > -4.0$ but moves to higher fields as the grain boundaries become more strongly normal (as $\alpha_{GB}$ decreases). Solid lines are fits to Eq. (71) with $r = 1.1$. Crosses represent comparison to typical experimental data for bronze route Nb$_3$Sn, taken from [1]. Inset: Fitting parameters for Eq. (71) as a function of $\alpha_{GB}$. (Lower): Maximum flux pinning force $F_p^{max}/J_D B_{c2}$ as a function of $\sqrt{1-\alpha_{GB}}$. Line fits are comparisons to Eq. (71) with $A = 0.25$, $r = 0.6$, $p = 0.5$ and $q = 2$, and to Eq. (68).

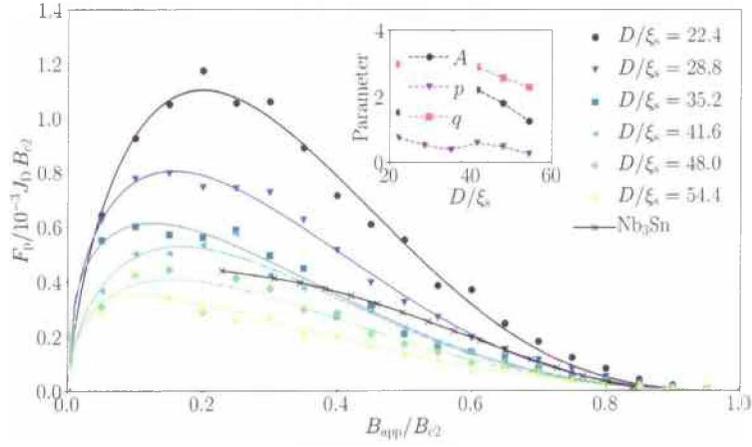

(a)

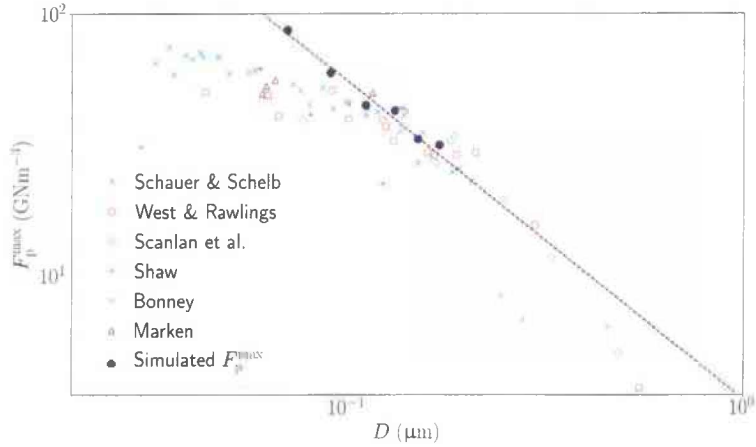

(b)

FIG. 11: (Upper): Normalised flux pinning force $F_\mathrm{p}/10^{-3}J_\mathrm{D}B_{c2}$ for a polycrystalline 3D system with varying mean grain size $D$. All other system parameters are set to the values given in Table I. Solid lines are fits to Eq. (71) with $r = 1.1$. Crosses represent comparison to typical experimental data for bronze route Nb$_3$Sn, taken from [1]. Inset: Critical current density $J_c$ as a function of applied field for varying grain size; colours correspond to main plot. (Lower): Maximum flux pinning force $F_\mathrm{p}^\mathrm{max}$ for the polycrystalline 3D system described in Table I with varying grain size $D$ compared to experimental data for the maximum flux pinning force measured in experimental Nb$_3$Sn samples taken from [63]. Dashed line represent fit to Eq. (71) with $p = 0.5$ and $q = 2$ with remaining free parameters found to be $A = 0.09$ and $r = 0.6$.

**TABLES**

| Parameter | Value |
|---|---|
| $h_{\{x,y,z\}}/\xi_s(T)$ | 0.5 |
| $L_x/\xi_s(T)$ | 150.0 |
| $L_y/\xi_s(T)$ | 150.0 |
| $L_z/\xi_s(T)$ | 150.0 |
| $D/\xi_s(T)$ | 22.4 |
| $d_{\text{GB}}/\xi_s(T)$ | 0.5 |
| $\alpha_{\text{GB}}$ | −2.0 |

TABLE I: Material parameters for the reference 3D polycrystalline system for 3D $J_c$ investigations. $J_c$ is decreased by 2.5% at each current step.